\documentstyle[12pt,epsf]{article}
\begin{document}
\def\bfi{{\bf\large 1}}
\def\Re{{\rm Re}}
\def\Im{{\rm Im}}
\def\bar{\overline}
\def\xcp{x_{cp}}
\def\fha{\hat{\cal A}}

\def\fa{{\cal A}} 
\def\fc{{\cal C}} 
\def\ff{{\cal F}} 
\def\fh{{\cal H}} 
\def\fm{{\cal M}} 
\def\fo{{\cal O}} 
\def\fr{{\cal R}} 
\def\ft{{\cal T}} 
\def\vu{v_u}
\def\vc{v_c}
\def\vt{v_t}
\newcommand{\BAR}{\overline}
\newcommand{\E}{\epsilon}
\newcommand{\ep}{{\eta^\prime}}
\newcommand{\etac}{{\eta_c}}
\newcommand{\te}{\theta_\eta}
\begin{flushright}
AMES-HET 97-10 \\
\end{flushright}
\bigskip
\begin{center}
{\large\bf The Possibility of
Large Direct CP Violation in $B\to K\pi$--Like Modes Due to
Long Distance Rescattering Effects and Implications for 
the Angle $\gamma$
}
\\
\vspace*{0.25 in}
%
%
%
%
David Atwood$^a$
and Amarjit Soni$^b$
\end{center}

\bigskip

\begin{flushleft}
$a$) Dept. of Physics and Astronomy, Iowa State University, Ames IA\ \
50010\\
$b$) Theory Group, Brookhaven National Laboratory, Upton, NY\ \ 11973
\end{flushleft}
\bigskip

\begin{quote} {\bf Abstract}: 
We consider the strong rescattering effects that can occur in the decays
such as $B\to K\pi$, $K^\ast\pi$, $K\rho\dots$ and their impact on direct
CP violation in these modes. First we discuss, in general, how the CPT
theorem constrains the resulting pattern of partial rate asymmetries
leading to different brands of direct CP\null. Traditional discussions
have centered around the absorptive part of the penguin graph which has
$\Delta I=0$ in $b\rightarrow s$ transitions and as a result causes
``simple'' CP violation; long-distance final state rescattering effects,
in general, will lead to a different pattern of CP: ``compound'' CP
violation. Predictions of simple CP are quite distinct from that of
compound CP\null. Final states rescattering phases in $B$ decays are
unlikely to be small possibly causing large compound CP violating partial
rate asymmetries in these modes. CPT theorem requires a cancellation of
PRA due to compound CP amongst the $K\pi$ states themselves; thus there
can be no net cancellation with other states such as $K^\ast\pi$, $K\rho$
etc.  Therefore, each class of such modes, namely $K \pi$, $K \rho$, $K^*
\pi$, $K a_1$ etc. can have large direct CP emanating from rescattering
effects.  Various repercussions for the angle $\gamma$ are also discussed. 
\end{quote}

\vspace*{0.5 in}

\section{Introduction}

Recent evidence from CLEO~\cite{cleoref} indicates that the long sought after
penguin dominated decays $B^0\to K^+\pi^-$ and  $B^+\to K^0\pi^+$ occur with
branching ratios  ($Br$) on the order of $10^{-5}$:

\begin{eqnarray}
Br(\bar B^0\to K^-\pi^+) &=&1.5{+0.5\atop -0.4}{+0.1\atop -0.1}\pm 
0.1 \times 10^{-5}       \nonumber\\
Br(B^+\to \bar K^0\pi^-) &=&2.3{+1.1\atop -1.0}{+0.2\atop -0.2}\pm 
0.2\times 10^{-5}
\label{cleonumbs}
\end{eqnarray}

\noindent where both modes have been averaged with their conjugates.

Using the short-distance (SD) Hamiltonian\cite{buchalla}, there has
been several recent theoretical calculations\cite{pengref} of such
exclusive modes. While these calculations are rather unreliable 
the relative
contributions to these processes from penguin and tree graphs
suggest that penguin operators, i.e.\ $b\to s g^*$,  will be
the dominant contributors. 
Nonetheless, tree
processes, i.e.\ $b\to W^* u\to u\bar u s$ could be an important feature of 
these decays through,
for example, interference effects with the penguin amplitudes. 

In this work we will primarily explore the possibility of relatively large 
direct CP violation driven by long distance (LD) 
rescattering effects in any of the following modes:

\begin{eqnarray}
B^-\to K^-\pi^0 & & B^-\to \bar K^0\pi^- \nonumber\\
\bar B^0\to K^-\pi^+ & & \bar B^0\to \bar K^0\pi^0. \label{kpilist}
\end{eqnarray}

\noindent In order to understand how CP violation will manifest itself in
these modes, we first prove general theorems that show how such large
effects may come about consistent with the constraints of CPT theorem. 
Applying this in the specific case of $K\pi$ final states, we shall see
that to have large partial rate asymmetries, there must be a significant
amount of inelastic rescattering with other light (e.g.\ $K+n\pi$)
states~\cite{donoghue}.  In particular, it is required that in the case of
$B^-$ decay the process $b\to u\bar u s$ contribute to the final state
$\bar K^0 \pi^-$ (which has a different quark content, i.e.\ $d\bar ds$). 
These strong rescattering effects may not in general be reliably
calculated but there are good reasons to believe that at the scale of the
$B$ mass their contributions are unlikely to be small.

CP violation emerging from these LD rescattering effects is rather
distinct from those governing the effects of penguin transitions, which
have been the focus of discussion over the past many years. In the latter
case the final states have $I=1/2$. The rescattering effects, on the
other hand, lead to states that are mixtures of $I=1/2$ and $3/2$. CPT
consideration along with unitarity of the $S$-matrix lead us to
categorize these as two brands of CP: simple CP and compound CP
violation. It is also useful to further subdivide simple CP into 
two categories: type I and type II, to be defined below.
The partial rate asymmetry cancellations in the various cases are
quite different with rather distinctive experimental predictions.

In particular, CP violation (i.e. simple CP-type I) driven by penguin
transitions~\cite{pertphase} (i.e. $\Delta I=0$) may be regarded as a
partial rate asymmetry of the quark level decay $b\to u\bar us$.  In this
case the partial rate asymmetry cancels with $b\to c\bar c s$.  At the
meson level this leads to a cancellation between $u\bar u s$ states such as
$K\pi$ and $c\bar c s$ states such as $D \bar D_s+n\pi$. On the other
hand, simple CP violation (type II) arising from rescattering effects, such as
in~\cite{donoghue}, lead to cancellations between mesonic states of light
quark content; for instance, $K\pi$ cancels against $K+n\pi$. Finally
compound CP violation is the result of interference between different
isospins and can only result in a cancellation of partial rate asymmetry
between different charge exchange modes, for example $B^-\to K^-\pi^0$
cancels against $B^-\to \bar K^0\pi^-$.

Using these interference effects we will then try to obtain information 
about the angle $\gamma$ of the CKM matrix. Bounds on $\gamma$ may be
deducible either assuming no LD rescattering effects or by assuming a
specific value for such effects.

It is important to note that while our discussion is largely in terms of 
$K\pi$, it applies more generally to similar states which involve only 
one amplitude. This would include all decays to a kaonic resonance and an 
isospin 1 meson where one of the two is a scalar or a pseudo-scalar, for 
instance:

\begin{eqnarray}
B\to K^* \pi \ \ \ K \rho   \ \ \ K(1^+) \pi \ \ \ K a_1\ \ \  
K(0^+) \rho\ \ \ \nonumber\\
K(0^+) \pi \ \ \ K a_2\ \ \ K^* a_0\ \ \ {\rm etc.} \label{allmodes}
\end{eqnarray}

\noindent In the case of compound CP violation, the CPT theorem dictates
that partial rate asymmetries arising from LD charge exchange
rescattering effects in the $K\pi$ system cannot cancel against those of
the $K^\ast \pi$ system (for example).  Thus each such class of final
state can independently have large direct CP emanating from long-distance 
rescattering effects. 


Note also that the PRA's from each of the three sources mentioned
in the above discussion are additive. This means that the net PRA
in some of the modes given above would be numerically bigger than 
that due only to compound CP, for example;
whereas in other cases it would be smaller, as a result
of partial cancellations.

Some of these modes have the useful property that the kaonic state is
self-tagging in the case where all the mesons are neutral.  For instance,
in the decay $\bar B^0\to \bar K^0\pi$ it is not possible to tell directly
whether a $B^0$ or a $\bar B^0$ initially decayed. However in the case
$\bar B^0\to \bar K^{0*} \pi$ followed by the subsequent decay $\bar
K^{0*}\to K^-\pi^+$ the final state could only come from a $\bar B^0$.

\section{Quark Level Processes}

In all of our subsequent discussion, we will be probing the same four 
quark level processes depicted in Fig.~1. Figs. 1(a)--1(c) represent 
penguin processes all of which mediate the decay 
$b\to u\bar u s$ ($d\bar d s$). Figure 
1(d) is a tree process which also gives $b\to u\bar u s$. 

Each of these quark level processes will take place in a meson
where the $b$-quark is combined with a light $u$ or $d$-quark to form a
$B^-$ or $\bar B^0$ meson. For definiteness, we take the final state to be
$K\pi$  so that we have one of the decays in eq.~(\ref{kpilist}) though
the generalization of our discussion to the modes in eq.~(\ref{allmodes})
is straightforward. 

Of primary concern to us is the possibility of CP violation due to the 
interference of these diagrams. Indeed in the phase convention of 
\cite{wolf_ckm} Figs.~1(c), 1(d) have a weak phase of $\gamma$ with 
respect to 1(a) or 1(b). In order to have CP violation manifested in the 
interference between these graphs however, it is also necessary that there
be  a strong rescattering phase. 

As pointed out in \cite{pertphase} one form of rescattering phase which
exists at the quark level arises in Fig.~1(b). In this case if the
invariant mass of the $u\bar u$ is larger than $2m_c$, the indicated cut
through the $c\bar c$ intermediate state will lead to an imaginary part for
this diagram.  The diagram in Fig.~1(c), which is the higher order contribution
to the rescattering in Fig.~1(d) will likewise generate an absorptive phase.
However, as discussed in the context of 
perturbation theory in~\cite{gerard}
and more generally  in \cite{wolfref}, the CPT theorem prevents the diagonal
rescattering $u\bar u\to u\bar u$ from contributing to CP violating asymmetries
at the quark level. In addition this particular higher order correction is  
likely to be numerically a small contribution to the amplitude.

In the paper \cite{palmer} a model for the contribution of the rescattering
effects of Fig.~1(b) to CP violation in $B\to K\pi$ is considered and it
is found that the resulting asymmetries are only a few percent,
although it must be kept in mind that these calculations have
significant uncertainties. In
\cite{gronau1} 
$K\pi$, $KK$ and $\pi\pi$ final states were considered in the context of
an SU(3) analysis which, as pointed out in \cite{gronau2,wolf2} implicitly
assumed that long distance rescattering of the final state was small. 
Specifically, in~\cite{gronau1} it was assumed that the tree graph,
Fig.~1(d) could not contribute to the decay $B^-\to \bar K^0 \pi^-$. 

If one accepts this assumption, there is an important implication
concerning the extraction of the CKM parameter $\gamma$ from experimental
data.  As is suggested in~\cite{wolf2}, the isospin amplitudes extracted
from the relations of \cite{gronau1} imply that given the total rates for
the decays in eq.~(\ref{kpilist}), even if one ignores CP violating
information (by adding each decay rate to its charge conjugate), one can
place a bound on the CP odd angle $\gamma$ which under some conditions may
be quite restrictive when compared with other experimental bounds on
$\gamma$. In \cite{man_fl} an interesting suggestion is made (as will be 
discussed below) that
with experimental data only slightly more precise than the current data
(\ref{cleonumbs}), an upper bound on the value of $\sin^2\gamma$ could be
established which would likely be fairly restrictive if the actual rates
are similar to the current central values.  Unfortunately this bound is
based on the assumption,
similar to \cite{gronau1} that 
either long distance rescattering effects are
not important or 
that all amplitudes are affected by such rescattering
according to a constant factor~\cite{man_fl_new}. We believe there is good
reason to think that this assumption may not hold.

Consider for instance the meson level Feynman diagrams in Fig.~2\cite{jeru}.
If one
naively calculates this diagram, due to the essentially massless meson 
exchange
in the $t$-channel, one obtains an answer which does not make
sense in the context of perturbation theory since contributions becomes so
large that perturbation theory is not trustworthy. In particular one obtains
the result that the loop contribution is larger than the initial $B\to
K\pi$ amplitude.  This would suggest that the LD rescattering phases are
unlikely to be small unless there are large cancellations.

In particular, as was also pointed out in \cite{wolf2}, in order for the
tree not to  contribute to the $\bar K^0\pi^-$ final state, there would have
to be a  remarkable coincidence of such long distance rescattering effects
adjusting both the isospins $I=1/2$ and $I=3/2$ by multiplying them by the 
same magnitude and phase. Unless the long distance rescattering phases are
vanishingly small it is highly implausible that such effects are 
independent of isospin.

Indeed, there is an analogous situation in $D$ decays~\cite{ddecay}
where a similar set of isospin amplitudes govern the Cabibbo allowed decay
$D\to K \pi$.  Here there are two isospin amplitudes $T_{3/2}$ and $T_{1/2}$
in terms of which the decay amplitudes are:

\begin{eqnarray}
A(D^+\to\pi^+\bar K^0)&=&\sqrt{3}T_{3/2} \nonumber\\
A(D^0\to\pi^+     K^-)&=&T_{3/2}+\sqrt{2}T_{1/2} \nonumber\\
A(D^0\to\pi^0\bar K^0)&=&\sqrt{2}T_{3/2}-T_{1/2}
\label{damps}
\end{eqnarray}

Using the branching ratios from \cite{pdb}, 
$\sqrt{\Gamma(D^+\to \pi^+\bar K^0)/\Gamma(D^0)}=0.104$,
$\sqrt{\Gamma(D^0\to \pi^+     K^-)/\Gamma(D^0)}=0.196$ and
$\sqrt{\Gamma(D^0\to \pi^0\bar K^0)/\Gamma(D^0)}=0.149$ from which one 
can solve for $|arg(T_{1/2}T_{3/2}^*)|\approx 86^\circ$\cite{lipkind}.

Another example of a $D^0$ decay which shows how long distance effects can
effectively annihilate one $q\bar q$ pair into another is $D^0\to K^0 \bar
K^0$ \cite{weakex}. In the usual singly Cabibbo suppressed charm decay $c\to
s\bar s u$, the final quark content of such a $D^0$ decay is $u\bar u s\bar
s$ which can only form $K^+K^-$ and so some rescattering must be involved
in the formation of the $K^0 \bar K^0$ state. The branching ratio

\begin{equation}
{Br(D^0\to K^0\bar K^0) \over Br(D^0\to K^+K^-)}\approx 0.3
\end{equation}

\noindent 
indicating such effects are prominent.
It is important to emphasize that the annihilation ($u\bar u\to
d\bar d$) in the above process is through long distance 
effects and cannot, in general  be calculated reliably 
through perturbation theory.
Therefore, in the analogous decay $B^-\to\bar K^0\pi^-$ we must seriously
consider the possibility that the tree graph has a substantial contribution
to the final state which cannot be estimated by short-distance perturbative
methods\cite{jeru}. 

One might hope that since the energies in the $B$ decay are much larger
than in the $D$, the rescattering effects through any given channel may be
greatly reduced. It has, however been suggested in \cite{donoghue} 
that there
are in fact many multi-body intermediate states which may contribute to the
rescattering process and they argue that the cumulative effect of all such
states does not decrease with $m_B$. 

The argument of~\cite{donoghue} is essentially as follows. First, from 
the optical theorem one can relate the forward scattering amplitude of 
$K\pi$ to the total cross section for $K\pi$:

\begin{equation}
Im({\cal M}_{K\pi\to K\pi}(s,t=0)) \approx s\sigma(K\pi)
\end{equation}

\noindent It is reasonable to assume that $\sigma(K\pi)$ follows the 
phenomenological scaling law of other total 
hadronic cross sections\cite{donoghue,landshoff,lipkins}:

\begin{equation}
\sigma(s)=X(s/s_0)^{0.08}+Y(s/s_0)^{-0.56} \label{eqseven}
\end{equation}

\noindent where $s_0\approx 1$ GeV\null. From this it follows that the above
imaginary part of the amplitude  scales like

\begin{equation}
Im({\cal M})\propto s^{1.08} \label{eqeight}
\end{equation}

If we now assume that the behavior of the amplitude ${\cal M}$ as a function
of $t$ for $t\leq 0$ is an exponential decrease ${\cal M}\propto
exp(-b|t|)$ 
(where $b\approx 0.25 GeV^-2$)
then it can be shown that the imaginary part of ${\cal M}
(B\to \pi K) \propto (M_b^2)^{0.08}$ just from integrating over the $K\pi$
intermediate states.  When taking into account a more detailed argument
involving Regge theory in the rescattering this is modified only slightly
to ${\cal M} (B\to \pi K) \propto (M_B^2)^{0.08}/\log(M_B^2/s_0)$. The
point is that there is little scaling with $M_B^2$. 

Since the amplitude for the elastic scattering is dominated by the
imaginary part of the amplitude, unitarity of the strong S-matrix can be
shown to imply that rescattering through other states, such as $K+n\pi$
will give an even bigger contribution to the strong phase of $B\to K\pi$
than the elastic rescattering channel. Thus the elastic rescattering will be
large and the inelastic rescattering \cite{inelastic} will be even
larger giving rise to a totally incalculable rescattering phase which
could well be appreciable even at the scale $(m_B)$ of the $B$ mass. 
Again, it is important to note that the large 
contributions to the 
rescattering amplitude assessed here 
cannot be estimated via perturbation theory. 

In the following sections, we consider the impact of such large phase 
shifts on the forms of CP violation which can occur in $B\to K\pi$. 
In particular the proportion of CP violation in pairs of charge exchange 
modes (e.g.\ $\bar K^0\pi^-$ versus $K^-\pi^0$) tells us about the nature 
of the rescattering processes involved.

First though, let us consider the implications of the CPT theorem in a very
general situation where some symmetry of the strong interaction is present.

\section{Implications of The CPT Theorem: Simple and Compound CP Violation}

The CPT theorem is an important prediction of relativistic quantum field
theory~\cite{textbook} and indeed all experimental information to date affirm
that CPT is an exact symmetry of  nature~\cite{textbook}. An important
consequence of this theorem is that the total decay rate of a particle, $A$ 
and its anti-particle, $\bar A$ are identical:

\begin{equation}
\Gamma(A)=\Gamma(\bar A)
\end{equation}

It does not however follow that the partial 
decay rate to a specific final state $\Gamma(A\to X)$ is the same as its 
CP conjugate $\Gamma(\bar A\to\bar X)$. In fact, defining 

\begin{equation}
\Delta\Gamma(A\to X) = \Gamma(A\to X)-\Gamma(\bar A\to \bar X)
\label{pradef}
\end{equation}

\noindent (where we will use $\Delta$ generally to mean the difference between
a quantity and its CP conjugate) if $\Delta\Gamma\neq 0$ CP is clearly
violated but CPT need not be.  CP violation of the form in
eq.~(\ref{pradef}) is referred to as a partial rate asymmetry. Clearly, if
CPT is not to be violated, all of the different partial rate
asymmetries \cite{notpra} present in a given decay must cancel. Thus there
must exist at least $n\geq 2$ states $\{X_1,\dots,X_n\}$ such that: 

\begin{equation}
\sum_{i=1}^{n}\ \Delta\Gamma(A\to X_i)=\Delta\Gamma(A)=0 \label{cpt_result}
\end{equation}

Let us start by considering the specific case of $B^-$ decay, i.e.\
$B^-\to K^-\pi^0$ and $B^-\to \bar K^0\pi^-$. As explained above, if there
is a partial rate asymmetry (PRA) in these modes,  they must exchange PRA
with some other state and  indeed, the state which it exchanges PRA with will 
depend fundamentally on the mechanism which gives rise to the PRA in the
first place. 

The state $B^-\to K^-\pi^0$ may exchange  partial rate asymmetry  in (at
least) two specific ways. First, there may be some net exchange 
of the two states $K^-\pi^0$ and $\bar K^0\pi^-$ with some other states;
and/or PRA in $K^-\pi^0$ may balance against the PRA in $\bar K^0\pi^-$.
We can characterize these two possibilities with the quantities:

\begin{eqnarray}
\Delta^+(B^-) & = & \Delta\Gamma(B^-\to K^-\pi^0) +
\Delta\Gamma(B^-\to \bar K^0\pi^-) \nonumber\\
\Delta^-(B^-) & = & \Delta\Gamma(B^-\to K^-\pi^0) - 
\Delta\Gamma(B^-\to \bar K^0\pi^-) \label{eqtwelve} 
\end{eqnarray}

\noindent Similarly, for the case of the $\bar B^0$ we have:

\begin{eqnarray}
\Delta^+(\bar B^0) & = &  \Delta\Gamma(B^-\to K^-\pi^+) +
\Delta\Gamma(B^-\to \bar K^0\pi^0) \nonumber\\
\Delta^-(\bar B^0) & = & \Delta\Gamma(B^-\to K^-\pi^+) -
\Delta\Gamma(B^-\to \bar K^0\pi^0) \label{eqthirteen}
\end{eqnarray}

\noindent We will refer to CP violating effects which cause $\Delta^+\neq
0$ as ``simple CP violation'' since, as we shall see, in this case the
exchange is between states of the same isospin while CP violation which
causes $\Delta^-\neq 0$ we will refer to as ``compound CP violation'' since it
can only result from the interference of two different isospin states.  CP
violation which maintains $\Delta^+=0$ is pure compound CP violation. In
general, however, it would be expected that both simple and compound CP
violation would be present.

There is a further distinction among the states which compensate for
$\Delta^+$, namely that portion which is exchanged with other final states
containing only light quarks and that portion which is exchanged with
states containing $c\bar c$ such as $D \bar D_s +n\pi$.  Let us define
$\Delta^+_{u\bar u}$ to be that portion of $\Delta^+$ exchanged with other
light quark states such as $K+n\pi$ and $\Delta^+_{c\bar c}$ to be that
portion exchanged with states containing a $c\bar c$. 
Thus, we can write

\begin{equation}
\Delta^+=\Delta^+_{c\bar c} + \Delta^+_{u\bar u} 
\end{equation}

\noindent For convenience, we are subdividing simple CP further.
The case $\Delta^+_{c\bar c} \neq 0$ is being dubbed type I
whereas $\Delta^+_{u\bar u} \neq 0$ is type II.  
The quantities $\Delta^+_{u\bar u}$ and $\Delta^+_{c\bar c}$ are
not, however separately experimentally observable. On the other hand, the
net exchange between all light quark states and all states containing
$c\bar c$ which we denote $\Delta^+(u\bar u)$ can be obtained from
experiment.  We can define this quantity by: 

\begin{equation}
\Delta^{+}(u\bar u) = \sum_i \Delta\Gamma( X_i^{u\bar u})
\end{equation}

\noindent where the sum is over all states $X_i^{u\bar u}$ which contain 
only  light
quarks. In effect, $\Delta^+(u\bar u)$ is the partial rate
asymmetry for $b\to  u\bar u s$ and so we expect it to correspond to the
perturbative calculation~\cite{pertphase} of the partial rate asymmetry
exchange between $b\to u\bar u s$ and $b\to c\bar c s$.

In any case, among the family of $K\pi$ final states (\ref{kpilist}), the
quantities $\Delta^+_{u\bar u}$ and $\Delta^-$ do not correspond in a
simple way to a quark level perturbative calculation. This is because in
terms of purely quark topologies there is no simple compensating process
for them provided by states consisting entirely of light quarks.  The long
distance rescattering effects which, from the discussion in the last
section, may be large, do provide such a mechanism. We will argue later
that such LD effects lead to large CP violation of the form
$\Delta^+_{u\bar u}$ and particularly $\Delta^-$ which may be as big as
$O(20\%)$ assuming $\gamma\simeq90^\circ$.

Let us now consider in very general terms some theorems which we
can apply to this case to show how the symmetries of the strong
interaction select which kinds of interference effects can contribute to
either $\Delta^+$ and $\Delta^-$. 

One way of understanding the CPT cancellation is to suppose that the  
Hamiltonian contains a strong piece which is CP invariant and a 
weak piece with terms with different complex phases (we will consider two
different such phases here for the purposes of  illustration):

\begin{equation}
\fh= \fh_s + \fh_{w1} e^{i\lambda_1} + \fh_{w2} e^{i\lambda_2} +h.c.
\end{equation}

\noindent In this formulation we refer to the strong force as the force
which predominate in the rescattering which in the cases we will be
interested in will be generated by QCD\null. The weak forces are those that
cause the initial decay and violate CP which in this case are
electro-weak interactions. All of the results we will consider will be to
lowest order in the weak interactions. 

Clearly we can therefore assume that the strong Hamiltonian does not contain
any  terms that allow the decay of $A\to X$. Now if $T$ is the weak transition
matrix and we expand it to first order in the weak Hamiltonian, 

\begin{eqnarray}
\fa(A\to X_i) & = & <X_i|T|A> = U_i e^{i\lambda_1}+V_i e^{i\lambda_2}
\nonumber\\
\fa(\bar A\to \bar X_i) & = & <\bar X_i|T|\bar A> = U_i e^{-i\lambda_1}+V_i
e^{-i\lambda_2} \label{general-phase}
\end{eqnarray}

\noindent where in general $U_i$ and $V_i$ are complex numbers and their
phases $\phi_U^i=arg(U_i)$ and $\phi_V^i=arg(V_i)$  are usually referred
to as strong phases since they may be regarded as being the result of rescattering
effects of the strong interaction.   Below, we will see in more detail how
the unitarity of the $S$ matrix  relates these phases to strong
effects. The partial rate asymmetry is thus  given (up to phase space factors)
by:  

\begin{equation}
\Delta\Gamma(B\to X_i) =-4|U_i|\
|V_i|\sin(\lambda_1-\lambda_2)\sin(\phi^i_U-\phi^i_V) 
\end{equation}

In~\cite{wolfref} it is shown that if there are only two states $X_1$ and
$X_2$ with partial rate asymmetries, the cancellation embodied in 
eq.~(\ref{cpt_result}) can be understood through the application of the
Cutkowski theorem. In each case the total strong phase results from the
rescattering through all possible intermediate states. It can, however be
shown that the part of the rescattering phase difference
$\phi_U^1-\phi_V^1$ which due to the contribution to $\phi_U^1$ which
results from rescattering through $X_2$ is equal and opposite to the
contribution to $\phi_V^2$ which result from rescattering through $X_1$. A
similar statement applies to the relation between contributions to
$\phi_V^2$ and $\phi_U^1$ and thus eq.~(\ref{cpt_result}) is realized.  If
there are more than two states, the contribution to the PRA of state $X_i$
resulting from rescattering through $X_j$ cancels 
the contribution to the partial rate asymmetry of $X_j$ resulting from the
rescattering through $X_i$ and thus the requirement of the CPT theorem is
affirmed. In this case we will say that $X_i$ exchanges PRA with $X_j$. 

Note that if there are more than 2 states, it cannot be experimentally
determined in detail how much PRA is exchanged between any given pair. For
instance, if there are 4 states, there are only 4 partial rate asymmetries
that may be observed but there are 6 possible pairs of states which may exchange
PRA\null.  

As the above description implies, it is a necessary condition that $X_i$
and $X_j$ can rescatter into each other for them to exchange PRA\null. In this
paper we wish to consider, in rather general terms, the role that the
symmetries of the strong interactions, play in the pattern of PRA 
exchanges between different final states.

Let us suppose that $R$ is a hermitian operator which commutes with the
strong Hamiltonian $\fh_s$ and that $R$ is invariant under CPT (i.e.\
$(CPT)$ \ $R^T$ \ $(CPT)^\dagger=R$).  Then the eigenspaces corresponding to
the various eigenvalues of $R$ will be invariant subspaces under
$\fh_s$. The decomposition of the various possible final states into the
eigenspaces of $R$ will allow us to understand which possible exchanges
of partial rate are allowed through the three theorems which we will
prove below:

\begin{quotation} {\it {\bf Theorem 1:} If $R$ is an operator invariant
under CPT and $[R,\fh_s]=0$ then, to first order in the weak interaction,
for each eigenvalue $r_i$ of $R$, $\sum_j \Delta\Gamma(A\to X_j)=0$
where the sum is taken over eigenstates of $R$ with eigenvalue $r_i$.}
\end{quotation}

{\bf Proof:} This theorem is a simple generalization of
eq.~(\ref{cpt_result}).  To prove it let us decompose following the
formalism of \cite{textbook} and write the $S$-matrix as follows: 

\begin{equation}
S=S_s +i T_W 
\end{equation}

\noindent Here $S_s$ is the strong rescattering matrix and is unitary 
and $S_s$ does not connect the initial state $A$ to the final states so
$\langle A|S_s|A \rangle=1$; $T_W$ is the first order transition matrix for
the weak interaction.  

If we apply unitarity to the above expression, we obtain the standard
result (where $\bfi$ is the identity matrix): 

\begin{equation}
\bfi =S^\dagger S =S_s^\dagger S_s  - i  T_W^\dagger S_s + i S_s^\dagger
T_W +  T_W^\dagger  T_W
\end{equation}

\noindent By unitarity $S^\dagger S=S_s^\dagger S_s=\bfi$ and so if we drop the
last term which is higher order in the weak interactions, we obtain

\begin{equation}
T_W^\dagger  =  S_s^\dagger  T_W  S_s^\dagger
\end{equation}

If we multiply this expression on the left by a final state $<X_i|$ and 
on the right by $|A>$ we obtain

\begin{equation}
<X_i|S_s^\dagger T_W  |A> =<X_i| T_W^\dagger |A>  \equiv <A|T_W|X_i>^*
\label{almost_there} 
\end{equation}

If we apply the CPT invariance to the right hand side of eq.~(\ref{almost_there})
and since by assumption $<A|S_s|A>=1$;  we obtain

\begin{equation}
<\bar X_i|T_W|\bar A>^* =<X_i| S_s^\dagger T_W | A>
\end{equation}

\noindent where the bar indicates the CPT transform of a given state, i.e.\
particles are transformed into their antiparticles with their spin degrees
of freedom reversed but momentum degrees of freedom the same. This
equation is identical to eq.~(1) of \cite{wolfref}.

Since $[R,H_s]=0$, $[R,S_s]=0$ and so if $r_i$ is an eigenvalue of $R$, the
space of eigenvectors  $\fr_i$ is an invariant subspace of $S_s$. In particular,
if $\Pi_i$ is the orthogonal projector onto $\fr_i$, $[\Pi_i,S_s]=0$. If
$\Gamma(A \to \fr_i)$ is the total decay rate of $A$ to states in  
$\fr_i$ then
 
\begin{eqnarray}
\Gamma(A\to \fr_i) &=& \sum_j  |<X_j |\Pi_i T_W | A> |^2 \nonumber\\
&=& <A|T_W^\dagger  \Pi_i T_W |A> \nonumber\\
&=& <A|T_W^\dagger    S_s S_s^\dagger \Pi_i T_W |A> \nonumber\\
&=& <A|T_W^\dagger    S_s \Pi_i S_s^\dagger  T_W |A> \label{forwards}
\end{eqnarray}

\noindent
where the sum over $j$ in the above indicates the sum over a complete set 
of states.

The corresponding decay rate for the antiparticle is 

\begin{eqnarray}
\Gamma(\bar A\to\fr_i ) &=& \sum_j |<\bar X_j| \Pi_i T_W |\bar A>|^2
\nonumber\\
&=& \sum_j |< X_j \Pi_i |  S_s^\dagger T_W | A>|^2 \nonumber\\
&=& <A| T_W^\dagger S_s \Pi_i S_s^\dagger T_W |A> \nonumber\\
&=&  \Gamma(A\to \fr_i) \label{backwards}
\end{eqnarray}

\noindent which are therefore identical hence $\Delta\Gamma(A\to \fr_i)=0$.

\noindent {\bf QED}
\medskip

This theorem is more specific than eq.~(\ref{cpt_result}) in that 
the PRA cancellations to first order in the weak interactions are shown to be 
between states that can rescatter into each other under the strong interaction.
In particular, states that are connected by the strong interactions must
share all possible quantum numbers preserved by $H_s$, for instance $r_i$
as above.  

Even if there is no PRA for any eigenstate of $R$, partial rate asymmetries
may still be present in states that are quantum 
mechanical mixtures of such eigenstates.  For these mixed states, we can
regard the PRA which will be present as being the
result of a separate mechanism due to the interference of the two
eigenstate channels. As the theorem below shows, this will result in a
distinctive pattern of net PRA exchange which we will
refer to as compound CP violation since two or more eigenstates of $R$
must be involved.  In general both simple and compound CP violation will
be present, however to understand what the features of compound CP
violation will be, let us consider the ideal case where no simple CP
violation is present, i.e.\ that for each eigenstate $X_i$ of $R$,
$\Delta\Gamma(A\to X_i)=0$. 

In this case, then, let $Y$ be a general state which
is a mixture of various eigenstates of $R$. Let us define $\ft(Y)$ to be
the smallest invariant subspace $R$ which includes $Y$. In particular,
$\ft(Y)$ will be spanned by $\{|Y>, R|Y>, R^2|Y>,\dots \}$ where the space
is exhausted after $n$ terms if $Y$ can be expressed as a linear
combination of $n$ eigenstates of $R$ ($n$ will be finite in all examples
we will consider). 

The partial rate asymmetries which may be present in such a case is however
restricted by the following theorem: 

\begin{quotation} {\it {\bf Theorem 2:}  Let $R$ be an operator invariant 
under CPT and $[R,\fh_s]=0$ and for all eigenstates of $R$, $X_i$, 
$\Delta\Gamma(A\to X_i)=0$. If $Y$ is a state which is not an eigenstate of
$R$ and $\Delta\Gamma(A\to Y)\neq 0$ then $Y$ has a net exchange of
partial rate asymmetry only with states in $\ft(Y)$ where $\ft(Y)$ is 
the smallest invariant subspace of $R$ which contains $Y$. Equivalently,
$\Delta\Gamma(A\to\ft(Y))=0$} 
\end{quotation}

{\bf Proof:} Let us denote by $\Pi_\ft$ the orthogonal projector onto the
subspace $\ft(Y)$. Since it is an invariant subspace of $R$,
$[R,\Pi_\ft]=0$ and thus for all $r_i$, $[\Pi_i,\Pi_\ft]=0$ and in fact
$\Pi_i\Pi_\ft$ is an orthogonal projector onto the subspace
$\ft(Y)\cap\fr_i$ which we will denote $\ft_i(Y)$. Let the states
$\{X_j^i\}$ be an orthonormal basis of $\ft_i(Y)$.  These states are
eigenstates of $R$ so by assumption none of these final states has a
partial rate asymmetry. Thus

\begin{equation}
\Delta\Gamma(A\to\ft(Y)) = \sum_i\sum_j \Delta\Gamma(A\to X_j^i) = 0
\end{equation}

\noindent {\bf QED}
\medskip

We can also restate this theorem in terms of the expectation  values for
observables of the final state. In particular if $\fo$ is some observable
we define the  expectation value  $<\fo>=<A|S^\dagger \fo S|A>$,
$\bar{<\fo>}=<A|S^\dagger \fo S|A>$ and $\Delta<\fo>=<\fo>-\bar{<\fo>}$.

\begin{quotation}
{\it {\bf Theorem 3:} If $\fo$ is a CPT invariant operator on the final
state where $[\fo,R]=0$ and for all eigenstates $X_i$ of $R$,  $\Delta
\Gamma(A\to X_i)=0$ then the expectation value $\Delta<\fo>=0$.}
\end{quotation}

{\bf Proof:}
This follows if we make an eigenstate decomposition in terms of states that
are both eigenstates of $R$ and of  $\fo$ (since $[R,\fo]=0$).  Using
the  CPT invariance of $\fo$, we can write this as:

\begin{equation}
\fo=\sum_k \lambda_k ( |X_k><X_k| + |\bar X_k><\bar X_k|)
\label{evexpandone}
\end{equation}

\noindent Since each of the above $X_i$ is an eigenstate of $R$, by assumption
the contribution  to $\Delta<\fo>$ of each of the terms in the above
expansion vanishes:

\begin{equation}
\Delta<\fo>=\sum_k \lambda_k  \Delta <( |X_k><X_k|)> =0
\label{evexpandtwo}
\end{equation}

\noindent {\bf QED}
\medskip

For our purposes, it is useful to state the above theorem in the 
logically equivalent form:

\begin{quotation}
{\it {\bf Corollary:} Let $\fo$ be a CPT invariant operator. if for all
eigenstates $X_i$ of $R$, $\Delta\Gamma(X_i)=0$, and $[R,H_S]=0$, then
$[\fo,R]\neq 0$ is a necessary condition for $\Delta<\fo>\neq 0$.}
\end{quotation}

\section{Examples of Compound CP Violation}

In summary we may restate the implication of the above theorems in terms
of what kinds of exchanges of PRA are possible. Thus we
have the following possibilities: 

\begin{enumerate} 

\item If $X_1$ is a state with a definite quantum number $r_i$ under some
symmetry $R$ which is conserved under the strong interaction, it can only
exchange PRA with a state $X_2$ which has the same quantum number $r_i$.
We will refer to this kind of PRA as being simple with respect to $R$. 

\item If no eigenstate $X_i$ of $R$ has a PRA, then a general state $Y$ can
only exchange PRA with other states related to it by the application of $R$.
We will refer to this kind of PRA as being compound with respect to $R$.

\item In general there may be both mechanisms of CP violation present. In
this case simple CP violation will account for the exchange of PRA between
$\ft(Y)$ and other states not in $\ft(Y)$ while compound CP violation can
only lead to exchanges within the states of $\ft(Y)$. 

\end{enumerate} 

\noindent Of course these are only necessary conditions for  the presence
of each kind of partial rate asymmetry; the specifics of the physical situation
will determine if either of these kinds of PRA will actually be present.
In this paper we wish to emphasize that even in the case 
of compound CP violation,
partial rate asymmetries can potentially be large. 

Before proceeding to our main example, decays of the form $B\to K\pi$, let
us consider, for instance, the CP violating effects discussed in
\cite{kpigamma}. Here decays such as $B^-\to \gamma K^*\pi$ and $B^-\to
\gamma K\pi$ were considered.  The analysis of the CP violation which is
discussed in \cite{kpigamma} may be facilitated by labelling which
instances are simple or compound with respect to the operator $R=J_h^2$.
Here $J_h$ denotes the total angular momentum of the hadronic part of the
final state. i.e.\ the $K\pi$ or $K^* \pi$ system. 

The model adopted in that paper is that the production of the final
hadronic final state is dominated by kaonic resonances.  In this model it
is possible that there is a PRA of specific angular momentum states which
would correspond to PRA of the form $\Delta\Gamma(B^-\to \gamma k_i)\neq
0$ for some kaonic resonance $k_i$.  For this to happen though, there must
be at least two states with the same $J^{PC}$ which exchange PRA, which is
possible for the two $1^{++}$ states considered, i.e. $K_1(1270)$
and $K_1(1400)$. This is an
explicit example of simple CP violation with respect to the symmetry operator
$J_h^2$

It was shown \cite{kpigamma}, however that this effect is likely to be quite
small.  In fact in the $K\pi$ case it cannot occur in this model since the
decay $1^+\to K\pi$ is forbidden. The larger effects which may occur in this
system are of the second type where we assume that for each specific
eigenstate of $J_h^2$, i.e.\ for each specific kaonic resonance channel, there
is no PRA but partial rate asymmetries result from the interference of two
such channels.  

To understand the type of CP violation which is present in the case of
$B^+\to \gamma K\pi$, let us define the angle $\theta$ between the $\pi$
and $\gamma$ in the $K\pi$ rest frame.  Clearly a final state with a
specific value of $\theta$ is, in general a mixture of all possible values
of $J_h$. Thus in accord with Theorem~2 even if no single angular momentum
state has a PRA, partial rate asymmetries may be
exchanged from one value of $\theta$ to another. The CP violation will
thus be manifested as a difference in the distribution in $\theta$ between
$B^+$ and $B^-$ decay even though the rate of decay integrated over
$\theta$ will be the same for both.  This would therefore be an example of
compound CP violation with respect to $J_h^2$. 

Let us now consider the case of primary interest, CP violation in the various
instances  of $B\to K\pi$ listed in eq.~(\ref{kpilist}). Here, the operator
which  provides the most useful characterization of various forms of CP 
violation is the total isospin, thus  we take $R=\vec I^2$. 

The final observed states (e.g.\  $\bar K^0\pi^-$ and $\bar K^-\pi^0$) are
not pure eigenstates of $\vec I^2$ but are linear combinations of $I=1/2$
and $I=3/2$ states. We will denote these eigenstates as $(K\pi)_{1/2}$ and
$(K\pi)_{3/2}$. 

Following  the discussion above, CP violations may cause partial rate
asymmetries among the states in (\ref{kpilist}) either as simple CP
violation in the $(K\pi)_{1/2}$ or $(K\pi)_{3/2}$ channels or compound CP
violation due to the interference of these channels. Simple CP violation
thus contributes to $\Delta^+$ while compound CP  violation contributes to
$\Delta^-$. 

Suppose that there is only compound CP violation so that $\Delta^+ = 0$ and 
$\Delta^- \neq 0$. This would clearly mean that the CP violation in, for 
instance, $K^-\pi^0$ is exactly compensated by the CP violation in 
$\bar K^0\pi^-$. It would be a mistake, however to jump to the conclusion
that the strong  phase involved in this CP violation $K^-\pi^0$ is due to
an intermediate  $\bar K^0 \pi^-$ state and vice versa. In general, if $n$
states are present there are  $n(n-1)/2$ instances where PRA exchange is
possible. If   $n\geq 3$, observing all $n$ partial rate asymmetries does not 
fix the $n(n-1)/2$ instances of exchanges. Thus compound CP violation in 
$K^-\pi^0$ could imply either that the only exchange is with  $\bar K^0\pi^-$
or that it has some net exchange with various other, perhaps multi-body states
while  $\bar  K^0\pi^-$ has an equal and opposite exchange with these or
similar  states. The argument in \cite{donoghue} together with Theorem 2
suggests  that the latter case would be the more likely scenario.

As we shall discuss below, it is unlikely that simple CP violation in
$(K\pi)_{3/2}$ will be large i.e.\ unless electroweak penguins are very
significant or physics beyond the Standard Model makes a large 
contribution to this decay. 
If simple CP violation is present in the $(K\pi)_{1/2}$ channel,
there are two possible kinds of states which PRA may be exchanged with, either
states which contain a $c\bar c$ pair such as $D \bar D_s+n\pi$ which we
will refer to as charm pair states or multi-body states that do not contain
a $c\bar c$ pair but only light quarks such as $K +n\pi$. 
The first of these is being called simple CP, type I, and the second
is type II.

In the case where the PRA is exchanged with $c\bar c$ states, it has been argued
\cite{pertphase} that the inclusive sum of the PRA exchange between all $c\bar
c$ states with all light quark final states may be estimated perturbatively
as the quark level PRA exchange between $b\to c\bar c s$ and $b\to
u\bar u s$. Models where this contribution to the PRA of $K\pi$ states has
been estimated through simple models of hadronization\cite{palmer} suggest
that it tends to be quite small i.e.\ O(a few \%) but with large
uncertainties. PRA exchange with multi-body
light quark states cannot  be calculated perturbatively. Furthermore, as
we have stressed, due to LD effects the rescattering phases involved in $B\to
K\pi$ can remain large  even at the high mass of the $B$ resulting in large
PRA's. 

If the LD rescattering phases are large, another important consequence which
we wish to emphasize in this paper is that compound CP violation in the modes
(\ref{kpilist}) is also likely to be large.  Following the reasoning in
\cite{donoghue} it seems that there is no simple argument which places an
a priori limit on the size of such effects. We will argue below that it is
not unreasonable  to have partial rate asymmetries on the order of
0(20\%) assuming $\sin\gamma\sim1$, though, again there is no reliable way
of calculating the phases that they depend on.

\section{CP Violation in $B\to K\pi$}

Let us now consider the physical mechanisms which may produce
these simple and compound CP violation in $B\to K\pi$.  To do this, we 
decompose  the amplitudes in terms of the weak phases of the theory as
in eq.~(\ref{general-phase}). Each amplitude is therefore expanded in terms
of elements of the Cabibbo Kobayashi Maskawa matrix as follows: 

\begin{eqnarray}
\fa &=& \vt \fa_t+ \vc \fa_c+ \vu \fa_u \nonumber\\
\bar \fa &=& \vt^* \fa_t+ \vc^* \fa_c+ \vu^* \fa_u \label{main_amp}
\end{eqnarray}

\noindent where

\begin{equation}
\vt=V_{tb} V_{ts}^* \ \ \ \  \vc=V_{cb} V_{cs}^* \ \ \ \  \vu=V_{ub}
V_{us}^* 
\end{equation}

\noindent In this expression, $\fa$ is the amplitude for a given $\bar
B$ meson decay and $\bar \fa$ is the amplitude for the charge conjugate
$B$ decay. As is well known, the equality of the factors $\fa_i$ in both
the $B$ and $\bar B$ amplitudes is a consequence of time reversal invariance
of the strong interaction.  Complex phases which are present in $\fa_i$
are the strong phases due to rescattering.  Because of the essential
nonperturbative origin of these rescattering phases it  is unlikely that
they may be accurately calculated. 

At the quark level, $\fa_t$ is generated by penguin graphs with an
internal $t$-quark as show in Fig.~1a and related higher order
corrections. Likewise $\fa_c$ is generated by graphs with an internal
$c$-quark as show in Fig.~1b. The amplitude $\fa_u$ may be generated
either by a penguin with a internal $u$-quark as show in Fig.~1c or by a
tree graph as shown in Fig.~1d. Indeed any distinction between these two
kinds of contributions is artificial since the penguin graph simply
represents the strong rescattering of the tree graph to light quark
($u\bar u$ or $d\bar d$) states.  In addition, unitarity of the CKM matrix
implies that $\vt + \vc + \vu =0$ therefore if we add an arbitrary constant
(i.e.\ independent of the quark mass) to each of the amplitudes
$\fa_i$, i.e.\ $\fa_i\to \fa_i+C$, the physics remains unaffected. Which
arbitrary constant one adds is purely a matter of convention; for our
purpose, we will choose to set $\fa_t\to 0$ so that we can write the
amplitudes as: 

\begin{eqnarray}
\fa &=& \vc\fha_c +\vu\fha_u \nonumber\\
\bar \fa &=& \vc^*\fha_c +\vu^* \fha_u \label{main_amphat}
\end{eqnarray}

\noindent where  $\fha_c=\fa_c-\fa_t$  and $\fha_u=\fa_u-\fa_t$. 

In the approximation of the CKM matrix used in \cite{wolf_ckm} the 
phase difference between the CKM angles in (\ref{main_amphat}) is

\begin{equation}
arg(\vu^* \vc) \approx arg(-V_{ud}V_{ub}^* V_{cd}^* V_{cb}) =\gamma
\end{equation}

\noindent where $\gamma$ is the phase of $\vu^*$ in this convention.
This CP odd angle may combine with a strong phase difference between
$\fha_c$ and $\fha_u$, resulting in CP violating effects proportional to
$\sin\gamma$.  

In particular, for each of the $B\to K\pi$ states let us define

\begin{equation}
\sigma={1\over 2}(Br + \bar{Br}) \ \ \ \  \delta={1\over 2}(Br - \bar{Br})
\ \ \ \ \xcp=\delta/\sigma
\end{equation}

\noindent where $Br$ is the branching ratio for the case involving the
$b$-quark while $\bar{Br}$ is the conjugate involving the anti-$b$-quark
and $\xcp$ is the partial rate asymmetry (PRA)
as it is traditionally defined~\cite{notpra}. In terms of the amplitudes 
above,
in units normalized to the total  branching ratio of the $B$ meson, 

\begin{eqnarray}
\sigma &=& |\vu\fha_u|^2 +|\vc\fha_c|^2 +2|\vu|\ |\vc|  
Re(\fha_u\fha_c^*)\cos(\gamma) \nonumber\\
\delta &=& +2|\vu|\ |\vc| Im(\fha_u\fha_c^*) \sin(\gamma).
\label{sig_del}
\end{eqnarray}

Let us now write these amplitudes in terms of their isospin components.
Here, one must realize that the penguin diagrams Fig.~1a, 1b and 1c are $\Delta
I=0$ transitions while the tree diagram Fig.~1d has both $\Delta I=0$ and
$\Delta I=1$ components.  The most general form of the amplitudes $\fha_c$
and $\fha_u$ allowed by  these isospin constraints are thus:

\begin{eqnarray}
\begin{array}{rclcrcl}
\fha_c(K^-\pi^0)&=&-A    
&\ & \fha_u(K^-\pi^0)&=&-B+\sqrt{2}D    \\                 
\fha_c(\bar K^0\pi^-)&=&
\sqrt{2}A &\ & \fha_u(\bar K^0\pi^-)&=&\sqrt{2}B+D \\      
\fha_c(K^-\pi^+)&=&-\sqrt{2}A &\ &  \fha_u(K^-\pi^+)&=&-\sqrt{2}C+D      \\ 
\fha_c(\bar K^0\pi^0)&=&A  &\ &  \fha_u(\bar K^0\pi^0)&=&C+\sqrt{2}D
\end{array}        \label{isospin_amps}
\end{eqnarray} 

\noindent Here each of $\{A$, $B$, $C$, $D\}$ is an amplitude which will
in general contain a strong phase, $\{A$, $B$, $C\}$ connect to $I=1/2$
final states of $K\pi$ and $D$ connects to the $I=3/2$ final state of
$K\pi$.  The assumption that $\rho= 0$ used in \cite{man_fl}
corresponds to

\begin{equation}
B=-D/\sqrt{2}. \label{manrel}
\end{equation}

\noindent As emphasized also in  \cite{wolf2,gronau2} 
this identity can only hold
if $B$ and $D$ have the same phase and, indeed, their magnitudes are required
to have the same ratio as would be the case in the absence of any large
rescattering effects.  

Of course the description in~\cite{donoghue} implies that this will not be
the case. On the other hand, in reference~\cite{man_fl_new} it is argued
that the penguin topologies are an accurate enough description of all
QCD rescattering effects in that the phase contributions from long distance
effects average out and such final state interaction effects only modify
the magnitude by a constant factor. In particular they suggest that $B^-\to 
\bar K^0\pi^-$ will
have its magnitude altered via long distance contributions but will still
receive no tree contribution as such. 

In view of the description in terms of the isospin amplitudes however, this
seems unlikely.  First of all, the rescattering implicit in Fig.~1(c) only
affects the $I=1/2$ amplitudes so if the effects were sizable in terms of
even only the magnitude then there would be a significant tree contribution
to $B^-\to \bar K^0\pi^-$. It is also hard to see how the 
phase shift in the $I=1/2$ and $I=3/2$ could be locked together. 
If indeed the phase shifts in the $I=1/2$ and $I=3/2$ fail to be the same 
as would be implied by eq.~(\ref{manrel}), 
that would in turn imply
at the very least,  
of compound CP violation. It seems therefore more likely that either 
rescattering effects in the $K\pi$ channel of $B$ decay are generally large,
in which case both simple and compound CP  violation would be present or
else  only  CP violation effects proportional to $\Delta_{c\bar c}$  are
present and the  description in \cite{gronau1,man_fl,man_fl_new} of the
decay $B^-\to \bar  K^0\pi^-$ is substantially correct.

If we denote the amplitudes for the decays as $m_1=\fa(K^-\pi^0)$;
$m_2=\fa(\bar K^0 \pi^-)$; $m_3=\fa(K^-\pi^+)$ and $m_4=\fa(\bar K^0\pi^0)$
then we can write the amplitudes as

\begin{eqnarray}
\begin{array}{ll}
m_1 = -A\vc -B\vu +\sqrt{2}D\vu & \bar m_1 = -A\vc^* -B\vu^*
+\sqrt{2}D\vu^* \\
m_2 = \sqrt{2}A\vc +\sqrt{2}B\vu +D\vu & \bar m_2 = \sqrt{2}A\vc^*
+\sqrt{2}B\vu^* +D\vu^*  \\
m_3 = -\sqrt{2}A\vc -\sqrt{2}C\vu +D\vu & \bar m_3 = -\sqrt{2}A\vc^*
-\sqrt{2}C\vu^* +D\vu^*  \\
m_4 = A\vc +C\vu +\sqrt{2}D\vu & \bar m_4 = A\vc^* +C\vu^*
+\sqrt{2}D\vu^*
\end{array}    \label{midef}    
\end{eqnarray}

Consider first the amplitudes $\{A$, $B$, $C\}$ \cite{isospin} which generate
the $K\pi$ states of $I=1/2$ which we will denote (for the $S=-1$ cases)
$(K\pi)_{1/2}^0$ and $(K\pi)_{1/2}^-$ (where the superscript indicates the
total charge).  Substituting into eq.~(\ref{sig_del}), we obtain  
the rates for decays to these states and their conjugates described by:

\begin{eqnarray}
{1\over 3}\sigma( (K\pi)^-_{1/2} )&=& |v_c|^2 | A|^2 + |v_u|^2 | B^2| 
+2 |v_u|\ | v_c|\ | A| \  | B|\cos\gamma\cos\phi_- \nonumber\\
{1\over 3}\delta( (K\pi)^-_{1/2} )&=& 
+2 | v_u|\ | v_c|\ | A|\ | B|\sin\gamma\sin\phi_-\nonumber\\
{1\over 3}\sigma( (K\pi)^0_{1/2} )&=& |v_c|^2  | A|^2 + |v_u|^2  | C^2| 
+2 | v_u|\ | v_c|\  |A|\  |C|\cos\gamma\cos\phi_0 \nonumber\\
{1\over 3}\delta( (K\pi)^0_{1/2} )&=& 
+2 | v_u|\ | v_c|\ | A|\ | C|\sin\gamma\sin\phi_0   \label{halfpra}
\end{eqnarray}

\noindent where $\phi_-=arg(BA^*)$ and $\phi_0=arg(CA^*)$. 

Clearly this effect is an example of simple CP violation. According to
Theorem~1 therefore, there must be an $I=1/2$ state which these states
exchange PRA with, whether they be $c\bar c$ states or light quark states.

In the case of $c\bar c$ states we can understand what is happening at the
quark level \cite{pertphase} from the schematic Feynman diagrams in Fig.~3a
and 3b.  In Fig.~3a we have a $c\bar c$ penguin contributing to
$\delta((K\pi)_{1/2})$ where the phase difference is generated by
the rescattering of $c\bar c$ through all possible on-shell states
indicated by the cut.  
This is simple CP-type I.
In fig.~3b we have the related diagram contributing
to $\delta(c\bar c s)$ through a $u\bar u$ penguin;  here the cut may
include, among other states, the $(K\pi)_{1/2}$ state. The contribution
that can be attributed to the $(K\pi)_{1/2}$ state is precisely the one
required to balance off the PRA in $(K\pi)_{1/2}$ final states. 



Let us turn our attention now to the case of an intermediate state which
is composed entirely of light quarks.  In this case we must consider all
states which have isospin $I=1/2$ (e.g. $K+n\pi$ or $K\eta^\prime+n\pi$
etc.).  An exchange of PRA in this case (simple CP-type II) 
will result from a difference,
$\phi_-$ and $\phi_0$ above, between the interaction phase of charm
penguin processes contributing to $A$ and the predominantly tree processes
contributing to $B$ and $C$. This can occur because the effective
hamiltonian for these two processes at the quark level has a different
dirac structure and so each process will couple differently to different
intermediate light quark states giving contributions to $\phi_-$ and
$\phi_0$.  Diagrammatically, this is shown in Fig.~3c where the hexagon
represents the contribution of the penguin operator to a multi-body
intermediate state including a kaon (e.g. $\bar K + n\pi$) and the circle
represents the tree contribution to the two body state $\bar K^0 \pi^-$.
Since this is simple CP violation, all of the states will be of the same
isospin, in this case $I=1/2$.  In Fig.~3(d) we show the compensating
process which gives an asymmetry to $B^- \to \bar K + n\pi$.  In the preceding
pages we have argued that these simple type II contributions may be large
or,  at least, are not bounded in any way.


For the $I=3/2$ final state there can be no simple PRA since the  penguin
diagrams produce only $I=1/2$ final states. Thus the total PRA summed over
all $K\pi$ final states is given by the $(K\pi)_{1/2}$ result above. In
particular, 

\begin{eqnarray}
\delta(K^-\pi^0)+\delta(\bar K^0\pi^-)&=&\delta((\bar K\pi)_{1/2}^-)
\nonumber\\
\delta(K^-\pi^+)+\delta(\bar K^0\pi^0)&=&\delta((\bar K\pi)_{1/2}^0)
\label{patterncpa}
\end{eqnarray}

Since the physical states that are actually detected are those in
eq.~(\ref{kpilist}) which are mixtures of the isospin eigenstates, compound
CP violation becomes possible giving $\Delta^-\neq0$.

To see this, consider as in Theorem~2, 
what happens in the limit of $\phi_-$, 
$\phi_0\to 0$ i.e.\ in the limit that there is no simple CP violation and
all of it is compound CP violation. In this case,

\begin{eqnarray}
\delta(\bar K^0\pi^-) &=& -\delta(K^-\pi^0) \nonumber\\
=
\delta(\bar K^0\pi^0) &=& -\delta(K^-\pi^+) \nonumber\\
&=& 
2\sqrt{2}|\vu|\ |\vc|\ | A|\ | D|  \sin\gamma \sin\Phi    
\label{patterncp} \end{eqnarray}

\noindent where $\Phi=arg(D A^*)$.  The equality in the first line and the
second line follow from Theorem~2 while the equality of all three lines follows
from the general  isospin considerations, in particular from the fact that
the penguin  process here is $\Delta I=0$, again ignoring the effects of
electroweak penguins. 

The isospin structure of the strong penguin also determines the pattern of
the simple CP violation given in eq.~(\ref{halfpra}).  In particular, if
{\it only simple CP violation} (i.e.\ $\Phi=0$) is present, these equations
become: 

\begin{eqnarray}
\delta(K^-\pi^0)={1\over 2}\delta(\bar K^0\pi^-)= 2||\vu\vc||\sin\gamma
\sin\phi_- \nonumber\\
\delta(\bar K^0\pi^0)={1\over 2}\delta( K^-\pi^+)= 2||\vu\vc||\sin\gamma
\sin\phi_0     \label{patterncpsimple}
\end{eqnarray}

Another way of expressing the pattern in eq.~(\ref{patterncpsimple}) is to
write it in terms of $\xcp$ where $\xcp(X_i)=\delta(X_i)/\sigma(X_i)$ i.e.
the PRA.  If we assume that the penguin processes (i.e. $\vc A$) dominates
$\sigma$ then

\begin{equation}
\xcp(K^-\pi^0)=\xcp(\bar K^0\pi^-)\ \ ; \ \  \xcp(\bar K^0\pi^0)=\xcp(
K^-\pi^+) 
\end{equation}

\noindent where if $B=C$ then all four values of $\xcp$ are equal.

If both simple and compound CP violation is present, from the combination
of eq.~(\ref{patterncpa}) and eq.~(\ref{patterncp}) we find that: 

\begin{equation}
2 \delta(K^-\pi^0) - \delta(\bar K^0\pi^-) - \delta(K^-\pi^+) +2
\delta(\bar K^0\pi^0) =0    \label{iso2}
\end{equation}

Since eq.~(\ref{iso2}) came from assuming that the penguin contribution is
$\Delta I=0$, a violation of this relation would imply that the light
quark pair is not made in a $I=0$ state.  If a penguin type process were
generating such a contribution this could mean that instead of being
produced via a virtual $g^*$ the quark anti-quark pair is produced via a
$\gamma^*$ or a $Z^*$ through either unexpectedly large electro-weak
penguin process or new physics penguins with large contributions.
Alternatively, tree processes involving perhaps extra $W$-bosons, charged
or neutral Higgs scalars could also lead to amplitudes with $\Delta I\neq
0$ that could violate eq.~(\ref{iso2}). 

In any case, in the context of the standard model, the important point to
note is that $\Phi$ is totally unconstrained by the CPT theorem. Furthermore
it is driven by LD rescattering effects in the $K\pi$ system so we cannot
say that it is small.

\section{Numerical Estimates}

Let us now estimate numerically the largest magnitude of CP violating
effects which might be present.  
For the purposes of this illustration, we will
consider the case where there is no simple CP violation but only compound 
CP violation $\phi_-=\phi_0=0$.
In order to obtain such a rough estimate
recall that the relations in eq.~(\ref{manrel}) would be true if final
state rescattering were turned off. If we assume as suggested
by~\cite{wolf2} that the main effect of such rescattering is to adjust
the respective isospin amplitudes by a phase, then the magnitudes, but
not the phases, obey the relation eq.~(\ref{manrel}), 
$|B| \approx |D|/\sqrt{2}$.
The largest CP violating effects would occur when $arg(A D^*)\approx
90^\circ$. Let us suppose that $B\approx C$ and define

\begin{equation}
r=|\vu\fha_u(K^-\pi^+)|/|\vc\fha_c(K^-\pi^+)|
\end{equation}

Thus, we find $||D\vu/(A\vc) ||\approx (\sqrt{2/5})r $ so that

\begin{eqnarray}
|\xcp(K^-\pi^0)|=|\xcp(\bar K^0\pi^0)| \nonumber\\
=2|\xcp(\bar K^0\pi^-)|=2|\xcp(K^-\pi^+)| \nonumber\\
\approx \sqrt{2} r \sin\gamma\sin\Phi \label{xcpphi}
\end{eqnarray}

\noindent Thus, if we suppose that $\Phi=\gamma=90^\circ$ then if $r=0.3$
\cite{london}, the above yields

\begin{eqnarray}
|\xcp(K^-\pi^0)|&=&|\xcp(\bar K^0\pi^0)|\approx0.42 
\nonumber\\
|\xcp(\bar K^0\pi^-)|&=&|\xcp(K^-\pi^+)|\approx 0.21.    \label{dphase}
\end{eqnarray}

As one can see, the isospin structure determines the pattern of CP 
violation as discussed in the last section. If the CP were simple and if
$B=C$,  $\xcp$ would be the same for each of the four modes assuming the
denominator is dominated by the penguin process.  On the other hand,
for the example of compound CP discussed above, PRA's in $K^-\pi^0$ and
$\bar K^0\pi^0$ mode are twice that in the $\bar K^0\pi^-$ and
$K^-\pi^+$ modes (see eq.~(\ref{xcpphi})).

\section{Bounding $\gamma$ From Experimental Data}

Let us now consider how information may be obtained about $\gamma$ through
the measurement of the rates of  $B\to K\pi$. 

First  let us consider  what may be learned about $\gamma$ from the two modes
that have actually been recently observed (\ref{cleonumbs}), namely $\bar
B^0\to K^-\pi^+$ and $B^-\to \bar K^0\pi^-$.  For each of these modes let
us define the following parameters which characterize the relative magnitudes
of various amplitudes: 

\begin{eqnarray}
r &=& ||\vu\fha_u(K^-\pi^+)||/||\vc\fha_c(K^-\pi^+)||   \nonumber\\
\rho &=& ||\fha_u(\bar K^0\pi^-)||/||\fha_u(K^-\pi^+)|| \nonumber\\
R &=& \sigma(K^+\pi^-)/\sigma(\bar K^0 \pi^-)     \label{mandefs}
\end{eqnarray}

In the paper \cite{man_fl} assuming $\rho=0$ it is shown that an accurate
measurement of $R$ may lead to a lower bound on $\cos\gamma$ especially if
information about $r$ from some other source is known. 

This bound comes about since, by isospin symmetry (the $u\bar u$
and $d\bar d$ pair from the gluon must be in a $I=0$ state),

\begin{equation}
\fha_c(K^-\pi^+)=\fha_c(\bar K^0 \pi^-)
\end{equation}

\noindent and since it is assumed that $\rho = 0$, 

\begin{equation}
\sigma(\bar K^0\pi^-)=|\vc\fha_c(K^-\pi^+)|^2
\end{equation}

\noindent Thus the observed ratio $R$ is given by

\begin{equation}
R=1+r^2+2 r \cos\gamma \cos\phi_-
\end{equation}

\noindent From the fact that $|\cos\phi_-|\leq 1$ we can in this
case infer that

\begin{equation}
|\cos\gamma|\geq  \left |{ 1+r^2-R \over 2 r }\right |   \label{manbound}
\end{equation}
 
Since this bound provides an upper bound on $\cos\gamma$, if $\gamma$ is
in the first or second quadrants (which is required by consistency with
CP violation in the $K^0_L$), there is some angle $\gamma_{max}$ such
that only $\gamma\leq\gamma_{max}$ and $\gamma\geq\pi-\gamma_{max}$ are
allowed. In Fig.~4 we show the allowed region for $\gamma$ in the first
quadrant as a function of $r$ given the values of $R=0.25$, $0.65$ and
$1.05$ (as shown by the solid curves).  The current experimental value is
$R=0.65\pm0.40$. From the graph it is clear that for the smaller values
of $R$ there is a lower bound on $\cos\gamma$ independent of any
information about $r$ corresponding to the peak in the curve. In fact if
$R<1$, then

\begin{equation}
\cos\gamma\geq\sqrt{1-R}.
\end{equation}

As pointed out in \cite{man_fl,london} one can argue that even the current
data from CLEO \cite{cleoref} would indicate an upper bound on $r$ if we
assume that $B^\pm\to\pi^\pm\pi^0$ is dominated by tree processes. If this
is true, then $SU(3)$ arguments would suggest that

\begin{equation}
r\approx { \lambda {f_K\over f_\pi} \sqrt{2 \sigma(\pi^-\pi^0)\over \sigma(
\bar K^0\pi^-)}   } 
\end{equation}

\noindent (where $\lambda=\theta_c$ is one of the CKM parameters from
\cite{wolf_ckm}). Thus given the current bound of
$\sigma(\pi^\pm\pi^0)<2\times 10^{-5}$ it follows that $r {< \atop \sim}  
0.5$. Factorization arguments in \cite{man_fl} suggest that $r\approx 0.2$
though this estimate has considerable uncertainty. 

\section{General Bound in The Presence of Long Distance Rescattering 
Effects}

It is probably unreasonable to assume that $\rho\to 0$. If 
however some argument or indirect evidence allows a bound on $\rho$ to be
known, $\rho\leq \rho_{max}$, then a bound  on $\cos\gamma$ may still be
obtained in some cases if $\rho_{max}\leq 1$. This is because

\begin{equation}
(1-r\rho_{max})^2\leq {\sigma(\bar  K^0\pi^-) \over ||\vc\fha_c(K^+\pi^-)||^2}
\leq (1+r\rho_{max})^2 
\end{equation}

\noindent so that

\begin{eqnarray}
\left | { 1+r^2-R(1+r\rho_{max} )^2 \over 2 r }\right | \leq |\cos\gamma|
&{\rm if}& 1+r^2\geq (1+r\rho_{max})^2 R \nonumber\\
\left | { 1+r^2-R(1-\rho_{max} r)^2 \over 2 r }\right | \leq |\cos\gamma|
&{\rm if}& 1+r^2\leq (1-r\rho_{max})^2 R    \label{boundB}
\end{eqnarray}

\noindent If $(1+r\rho_{max})^2 R > 1+r^2 > (1-r\rho_{max})^2 R$
then there is no bound on $\cos\gamma$. If $\rho_{max}=0.3$ the bounds 
for various values of R are shown in Fig.~5 with dashed lines.

There is some prospect of obtaining information about the value of $\rho$
through the study of the analogous process $B^0\to K^+ K^-$. In this case
neither a tree decay nor a penguin decay may lead to the final state quark
content $u\bar u s \bar s$. The tree decay $b\to u\bar u d$ can, however
produce, for instance, a $\pi\pi$ state that can rescatter to $K^+K^-$ and
likewise a penguin decay $b\to s\bar s d$, $u\bar u d$, and $d\bar d d$
can lead to a $\pi\pi$ or $K^0 \bar K^0$ state which may rescatter to
$K^+K^-$. Thus, by comparing the rate of $B^0\to K^+K^-$ to $K^0\bar K^0$,
$\pi^+\pi^-$ or $\pi^0\pi^0$, it may be possible to put a bound on $\rho$.
In particular if $B^0\to K^+K^-$ is much smaller than the other processes
then the assumption of \cite{man_fl,man_fl_new} would be vindicated.

\section{Constraints on $\gamma$ via Direct CP Violation in $B\to K\pi$}

In view of the fact that it may not be possible to derive a bound on
$\rho$, it would be useful to have another way to find a bound on
$\gamma$. If CP violation is discovered in any of the four modes $B\to
K\pi$ (i.e. $\delta\neq 0$) then a lower bound can be placed on
$\sin\gamma$. 

To understand how this works, suppose that $\gamma$ and $r$ were known.
Then, the system of equations
eq.~(\ref{sig_del})
can be solved for a positive real value of $|\vu\fha_u|$ if and only if

\begin{equation}
|\sin\gamma| \geq {\xcp\over 2\sqrt{ 1-\xcp^2} }
\left | { 1-r^2\over r }\right |      \label{boundC}
\end{equation}

\noindent where $\xcp=\delta/\sigma$. Thus, if $\gamma$ is in the first or
second quadrant this bound will mean that there is a value of
$\gamma_{min}$ such that only $\gamma_{min}\leq \gamma \leq
\pi-\gamma_{min}$ is allowed as a function of $r$.  In Fig.~5 we show this
bound as a function of $r$ in the first quadrant.  

From eq.~(\ref{boundC}) [see Fig.~5] if $r=1$ there is no lower bound on 
$|\sin\gamma|$. This corresponds to a situation where the penguin and tree
happen to almost exactly cancel so that a small value  of $\gamma$
is amplified to a large value of $\delta$ due to almost total destructive
interference. Since $r$ is likely to be smaller than 1, this singular
configuration is probably not a problem, future experimental measurement
of $\pi\pi$ modes together with $SU(3)$ arguments should help to clarify
what a reasonable value of $r$ is.  If an overall upper bound on the value
of $r$, $r\leq r_{max}\leq 1$ is known, then the lower bound on
$|\sin\gamma|$ for all values of $r\leq r_{min}$ will be obtained by
substituting $r_{min}$ into eq.~(\ref{boundC}).  A similar statement is
true if a lower bound on $r\geq r_{max}\geq 1$ is known. 

For instance, as a numerical example, if it were true that the restriction on
$r$ of $r_{max}=0.2$ can be obtained, a value of $\xcp=0.3$ would lead to
the bound $50^\circ\leq\gamma\leq 130^\circ$ while if $\xcp=.1$ gives
$14^\circ\leq\gamma\leq 176^\circ$. One can see that to put bounds on
$\gamma$ that are interesting from the perspective of the Standard Model,
one must have an instance of $\xcp\geq O(0.1)$ for at least one of the 
modes.

\section{Extracting Information about $\gamma$ from 
Direct CP in $B\to K\pi$
-like Modes.}

Let us now consider the case where full experimental information about 
this system (\ref{kpilist}) is available.
If  all four branching
ratios and their conjugates may be observed, it is still not in general
possible to solve for $\gamma$ without making some additional assumption. 
One can, however, obtain the combination: 

\begin{equation}
Q=|\vc A|\sin\gamma
\end{equation}

\noindent The experimental determination of the branching ratios for each
of the  four modes and their conjugates allows us to determine $|m_i|$ and
$|\bar m_i|$ of eq.~{\ref{midef}}. i.e.\ eight quantities in all
subject to one constraint i.e.\ eq.~(\ref{iso2}).

We can most easily obtain information about the amplitudes from the
observable quantities by noting that a common strong phase ($\phi_D$)
is not soluble and by rewriting eq.~(\ref{midef}) in terms of the expressions

\begin{eqnarray}
f         &=& 3  e^{-i(-\gamma+\phi_D)} \vu D   \nonumber\\
g_1       &=& 3  e^{-i(-\gamma+\phi_D)}(\vu B+\vc A)   \nonumber\\
\bar g_1  &=& 3  e^{-i(+\gamma+\phi_D)} (\vu^* B+\vc^* A)   \nonumber\\
g_2       &=&-3  e^{-i(-\gamma+\phi_D)}  (\vu C+\vc A)   \nonumber\\
\bar g_2  &=&-3  e^{-i(+\gamma+\phi_D)}  (\vu^* C+\vc^* A) \label{fg12}
\end{eqnarray}

\noindent where $\phi_D=arg(D)$ so $f$ is real and $g_i$ and $\bar g_i$ 
are general complex numbers which satisfy 

\begin{equation}
g_1+g_2-\bar g_1-\bar g_2=0
\end{equation}

We then obtain:

\begin{eqnarray}
\begin{array}{ll}
3|m_1| = |-g_1+\sqrt{2} f| & 
3|\bar m_1| = |-\bar g_1+\sqrt{2} f| \\ 
3|m_2| = |\sqrt{2} g_1+ f|  &
3|\bar m_2| = |\sqrt{2} \bar g_1+ f| \\
3|m_3| = |\sqrt{2} g_2+ f| & 
3|\bar m_3| = |\sqrt{2} \bar g_2+ f| \\ 
3|m_4| = |-g_2+\sqrt{2} f| &
3|\bar m_4| = |-\bar g_2+\sqrt{2} f| \\
\end{array}
\end{eqnarray}

These equations may be solved to obtain the complex values of 
$g_i$, $\bar g_i$ as well as the real number $f$, though the solutions 
will have some discrete ambiguities since they require the solution to 
polynomial equations.

The quantity $Q$ may thus be expressed as 

\begin{equation}
Q=|\vc A|\sin\gamma=|g_1-\bar g_1|/6
\end{equation}

Furthermore, from $g_1$ and $\bar g_1$  we may also discover if there is
indeed a strong phase difference $\Phi=arg(DA^*)$ because

\begin{equation}
\Phi=arg(i(g_1-\bar g_1))
\end{equation}

In addition we can learn the phase of $|B-C|$ since 

\begin{equation}
{1\over 3}|g_1+g_2|=(B-C)|\vu|
\end{equation}

The simple point is that there are seven independent quantities that are
measured 
since the eight values of $|m_i|$ and $|\bar m_i|$ are subject to the 
constraint eq.~(\ref{iso2}). On the other hand, 
the right hand side of eq.~(\ref{fg12}) 
depend on 
eight unknowns:
$\gamma$,
$|\vc|Re(A)$,
$|\vc|Im(A)$,
$|\vu|Re(B)$,
$|\vu|Im(B)$,
$|\vu|Re(C)$ 
and
$|\vu|Im(C)$
(note that the observables do not depend on an overall strong phase, here 
taken as $\phi_D$).
Thus 
$\gamma$ cannot be determined from these equations.

However, if we know the value of $\rho$ we may obtain the
ratio 

\begin{equation}
r_B=|B/D|= \left |  {\sqrt{2}\rho^2\pm3\rho+\sqrt{2}\over 1-2\rho^2} 
\right |
\end{equation}

\noindent where the $\pm$ in the above represents a two fold ambiguity. 
From this 

\begin{equation}
\gamma=arg( (1-i\lambda)g_1 - (1-i\lambda)\bar g_1)
\end{equation}

\noindent where $\lambda$ is one of the two solutions to 

\begin{equation}
|(1+i\lambda)g_1 + (1-i\lambda)\bar g_1|=2f r_B.
\end{equation}

In the above we assume that the decays of $B^0$ are self tagging and so
oscillation effects need not be taken into account.  This would not be
true for $\bar B^0\to\bar K^0\pi^0$, however in the analogous case where
the $K^0$ is replaced with the $K^{0*}$ which decays
to a charged $K^\pm$ the decay chain will be self tagging. Thus
$\delta(\bar K^{*0}\pi^0)$ may be determined through the comparison of
the decay chain $\bar B^0 \to \bar K^{*0}\pi^0 \to  K^{-}\pi^+\pi^0$
to $B^0 \to K^{*0}\pi^0 \to K^{+}\pi^-\pi^0$.

In the case where the decay is not self tagging such as $\bar B^0\to \bar
K^0\pi^0$ or $\bar B^0\to \bar K^0\rho^0$, we can still carry out the
analysis through the use of eq.~(\ref{iso2}). Consider for instance the
case $\bar B^0\to \bar K^0\pi^0$. In this case oscillation effects will
not alter the observed value of $\sigma(\bar K^0\pi^0)$ while $\delta(\bar
K^0\pi^0)$ may be obtained through eq.~(\ref{iso2}).

Of course using this equation assumes the isospin structure due to the
presence on the quark level of only the tree and strong penguin diagrams.
In order to confirm this one can independently check the value of
$\delta(\bar K^0\pi^0)$ by factoring in the oscillation effects. Let us
consider the experimental situation as it exists at an $e^+e^-$ collider
where a $B^0\bar B^0$ pair is produced, one of the pair undergoes a
tagging decay and the other one decays (for instance) to $K_s\pi^0$.  Here
we will consider the situation where the times of the decay cannot be
determined (as would likely be true for $K_s\pi^0$) and so we consider
only time integrated quantities.

Let us denote a tagging decay that indicates a $B^0$ meson (such as
$e^+\nu D^-$) by $B^0\to tag$ and a tagging decay that indicates a $\bar
B^0$ meson (such as $e^-\bar \nu D^+$) by $\bar B^0\to \bar{tag}$. If we
designate the neutral $B$-meson that undergoes the tagging decay as $B_1$
and the neutral $B$ meson which undergoes the decay to $K\pi$ as $B_2$
then we can define the following observable time integrated quantities:

\begin{eqnarray}
{1\over 2} \hat\sigma(K_s\pi^0)&=&{1\over 2}( Br(\bar B^0\to K_s\pi^0) +
Br(B^0\to K_s\pi^0))
\nonumber\\
{1\over 2} \hat\delta(K_s\pi^0) &=& { Br(B_1\to{tag};B_2\to K_s\pi^0) -
Br(B_1\to\bar{tag};B_2\to K_s\pi^0) \over Br(B_1\to {tag}) + Br(B_1\to
\bar{tag})}
\nonumber\\
\ \ \ \ 
\end{eqnarray}

\noindent
These may be related to $\sigma$ and $\delta$ via:

\begin{eqnarray}
\hat\sigma(K_s\pi^0)&=&
\sigma(\bar K^0\pi^0)
\nonumber\\
\hat\delta(K_s\pi^0)&=&
{1\over 1+x_d^2}
\delta(\bar K^0\pi^0)
\label{dilution}
\end{eqnarray}

\noindent
where $x_d=\Delta m_B/\Gamma_B$.

Thus if $\hat \sigma(K_s\pi^0)$ and $\hat\delta(K_s\pi^0)$ are observed
experimentally, the quantities $\sigma(\bar K^0\pi^0)$ and $\delta(\bar
K^0\pi^0)$ may be found from eq.~(\ref{dilution}) which gives us $|m_4|$
and $|\bar m_4|$. The analysis for extracting $Q$ then proceeds as given
above. For the $B^0$, the experimental value for $x_d$ is about $.73$
hence the factor $1/(1+x_d^2)$ in eq.~(\ref{dilution}) is about $.65$.

\section{Several Shots at Large Direct (Compound) CP.}

It is important to understand that due to Theorem 2, the partial rate
asymmetries in $B\to K\pi$ that are driven by LD rescattering effects
leading to compound CP cannot cancel with similar PRA's in $B\to
K^\ast\pi$ system, for instance. Since, as a rule, we should anticipate LD
effects to cause possibly large, unpredictable, phases in all such modes
(see eq.~(\ref{allmodes})) therefore experimentally we get several
independent shots at the consequences of large direct CP by searching for
all of these modes. We note, in passing, in this context that large final
state rescattering phases have been seen in $D\to K\pi$, $K^\ast\pi$ 
and in $K\rho$\cite{ddecay}. 

\section{Conclusions}

Traditional discussions of direct CP in $B$ decays \cite{palmer,kramertwo} 
have
been centered around that emerging from the absorptive part of the penguin
graph \cite{pertphase}. We are labelling this ``simple CP violation''
as, for $b \rightarrow s$ transitions,
it involves $\Delta I=0$ effective interaction only. 
Simple CP of type I entails partial width cancellation against
$c\bar c$ states whereas for type II the cancellation is 
with light quark states which contribute through final
state interactions\cite{donoghue}. 
Long distance
rescattering effects can cause another brand of CP
violation---``compound CP violation'' involving mixtures of eigenstates 
of isospin.
We have discussed CPT
constraints governing the PRA's in the various cases. In particular, the
pattern of asymmetries in $B\to K\pi$ modes in these cases is quite
different.

We have also examined the repercussions of the long-distance rescattering
effects for constraints on the CKM angle $\gamma$.  Since at $m_B$ LD
rescattering effects in $B\to K\pi$-like modes are unlikely to be small
they need to be taken into account.  Full experimental information in the
$K\pi$ helps in deducing useful constraints on $\gamma$.

Since PRA due to compound CP in $B\to K\pi$ cannot cancel with those
(say) in $B\to K\rho$, each class of these final states would exhibit
PRA dictated by the corresponding rescattering effects in the
respective channel.


PRA's from different sources of CP, discussed herein, are additive.
Thus in some of these modes the net PRA will be bigger than
that only due to compound CP, for example; in other cases, due
to partial cancellations, it could be smaller.

\newpage
\bigskip

\noindent {\large\bf Acknowledgement}
\medskip

This research was supported in part by DOE contracts DE-AC02-76CH00016
(BNL) and DE-FG02-94ER40817 (ISU).

\bigskip
\noindent {\large\bf Note Added:}
\medskip

In the final stages of preparation of this paper we became aware of two 
preprints which discuss some of the same issues as this paper.
These are: M.~Neubert, hep-ph/9712224 and 
A.~Falk, A.~Kagan, Y.~Nir and A.~Petrov, hep-ph/9712225.

\newpage
\begin{center}
{\large\bf Figure Captions}
\end{center}

\begin{itemize}

\item[Figure 1] Lowest order Feynman diagrams for various quark level
processes which lead to $B\to K\pi$:  (a) A penguin diagram with an
intermediate $t$-quark (b) A penguin diagram with an intermediate
$c$-quark (c) A penguin diagram with an intermediate $u$-quark (d) A tree
graph $b\to u\bar u s$.  In graphs (c) and (d) cuts are shown where there
can be intermediate on-shell states. 

\item[Figure 2]
Examples of meson level diagrams for the rescattering of a $K^-\pi^0$ 
final state  to the final state $\pi^-\bar K^0$. 


\item[Figure 3] Quark level Feynman diagrams that contribute to the
partial rate asymmetry for for simple CP violation of type I at the quark
level, decays involving $b\to u\bar u s$ and $b\to c\bar c s$, and CP
violation of type II at the meson level. Figure 3(a) shows a penguin
contribution which generates a partial rate asymmetry in $(K\pi)_{1/2}$
which is simple CP violation of type I through the interference with the
tree diagram where the strong phase of the penguin is generated by the
$c\bar c$ cut indicated.  Figure 3(b) shows a contribution to the partial
rate asymmetry for $b\to c \bar c s$ through the interference of a penguin
with an internal $u$-quark and the tree. The cut here includes $K\pi$
states and the contribution of those $K\pi$ states will be exactly
opposite to the partial rate asymmetry of $K\pi$ in 3a.  Figure 3(c) shows
a contribution to simple CP violation of type II where the hexagon
indicates a penguin process, the circle indicates a tree process and the
box indicates strong rescattering. In this case the intermediate state is
a multi-body e.g. $K+n\pi$. Figure 3(d) shows the process which compensates
for the partial rate asymmetry in 3(c).



\item[Figure 4] An example of the bounds that may be obtained on $\gamma$
from the observation of $\sigma(K^+\pi^-)$ and $\sigma(\pi^-\bar K^0)$
under various assumptions as a function of $r$.  In all cases the allowed
region is below the curve. The solid curves correspond to the case
$\rho=0$ for the values of $R=0.25$, $0.65$ and $1.05$ as indicated. The
dashed curve is the bound given $R=0.65$ and $\rho_{max}=0.3$; the dotted
curve is for $R=0.65$ and $\rho_{max}=0.5$ while the dot-dashed curve is
for $R=0.65$ and $\rho_{max}=1$.

\item[Figure 5] The bounds that may be obtained on $\gamma$ from the
observation of $x_{cp}$ for some mode $B\to K\pi$. Here the allowed region
is above the curves. The bound if $x_{cp}=0.03$ is shown in the dashed
curve, the bound if $x_{cp}=0.1$ is shown in the solid curve, and the
bound if $x_{cp}=0.3$ is shown in the dot-dashed curve. 
\end{itemize}

\newpage

\begin{figure}[ht]
\hspace*{-0.5in}
\vspace*{0in}
\epsfysize 7 in
\mbox{\epsfbox{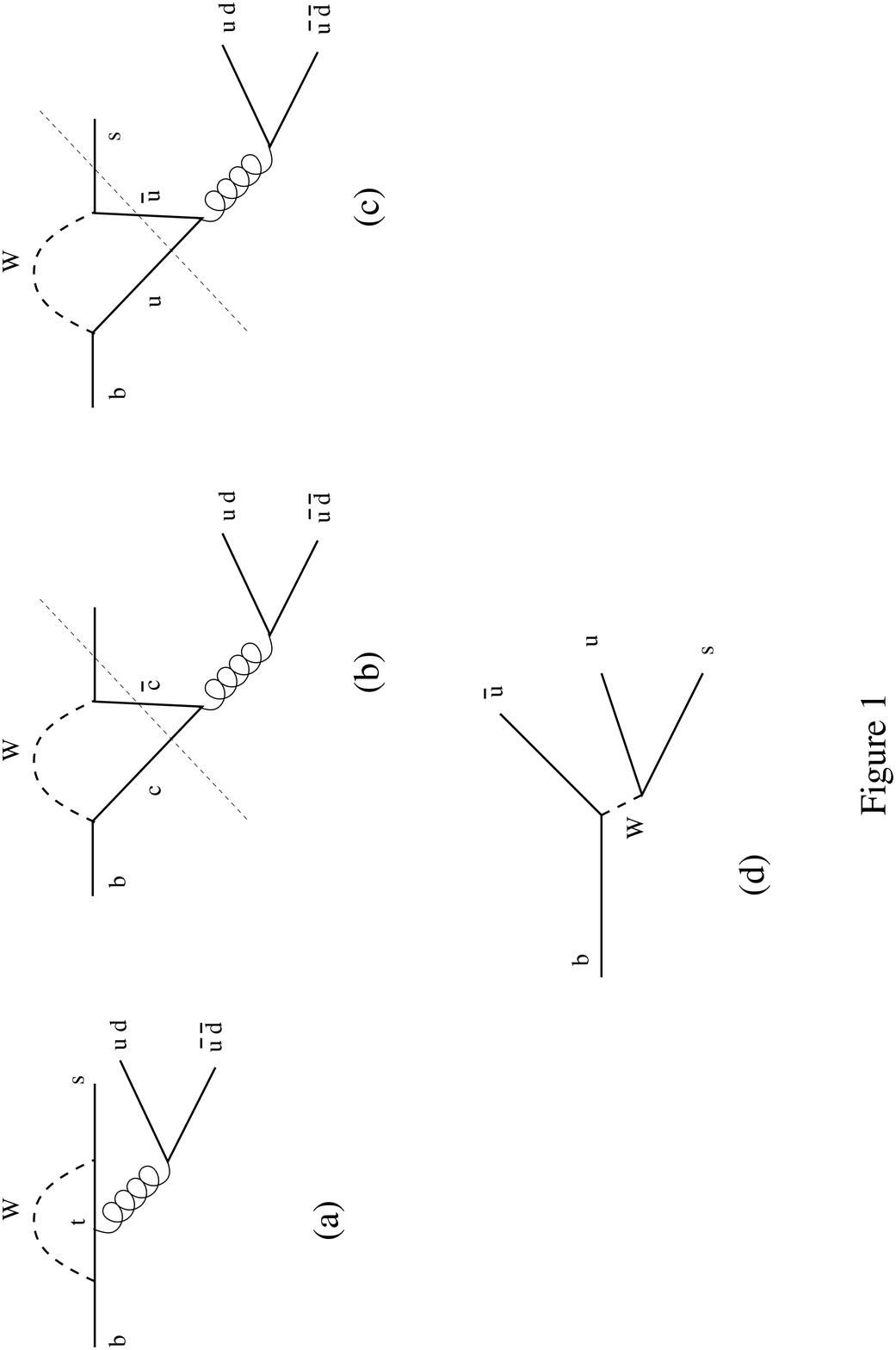}}
\end{figure}

\newpage

\begin{figure}[ht]
\hspace*{0 in}
\vspace*{ 0 in}
\epsfysize 5 in
\mbox{\epsfbox{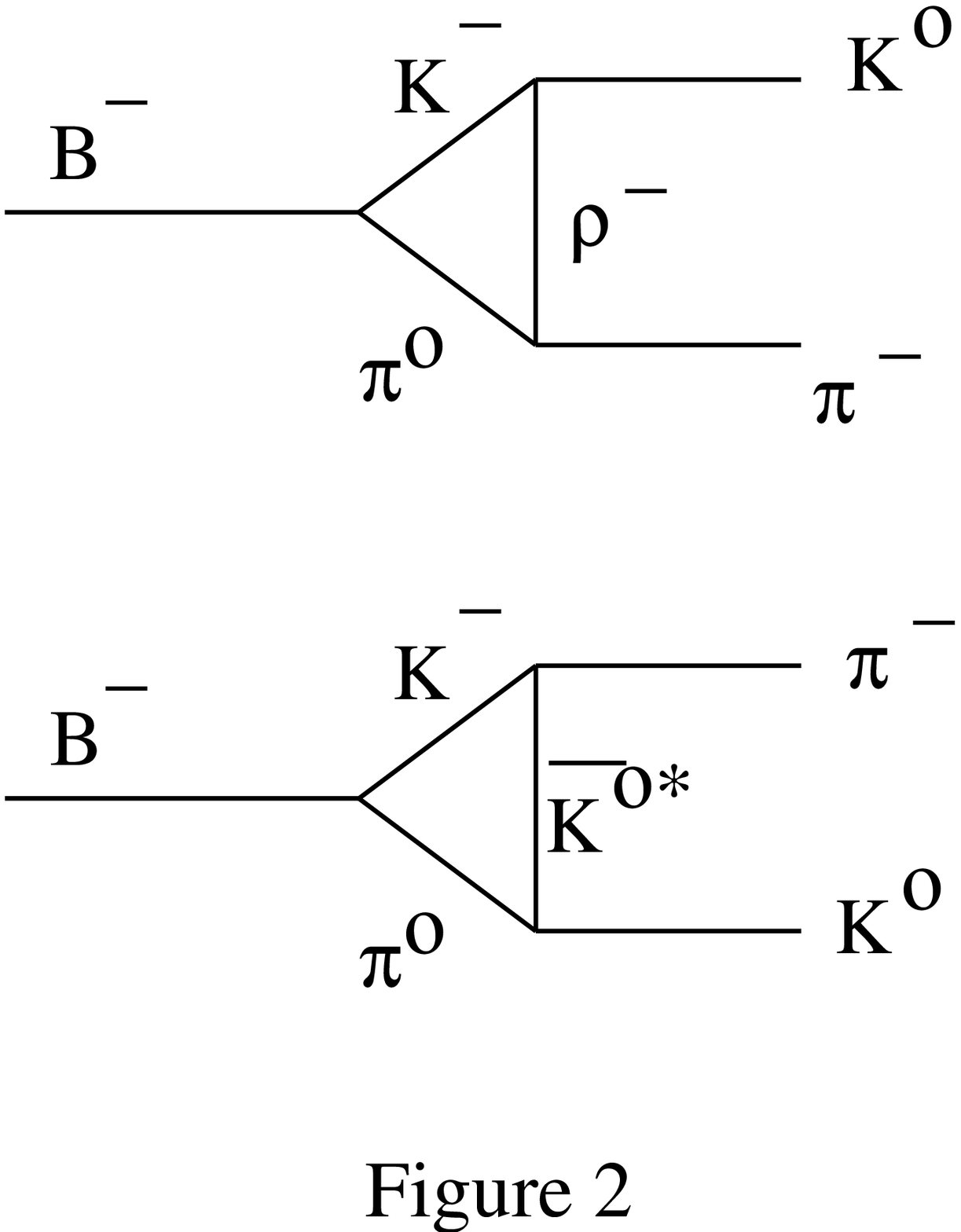}}
\end{figure}

\newpage

\begin{figure}[ht]
\hspace*{-0.5in}
\vspace*{0in}
\epsfysize 7 in
\mbox{\epsfbox{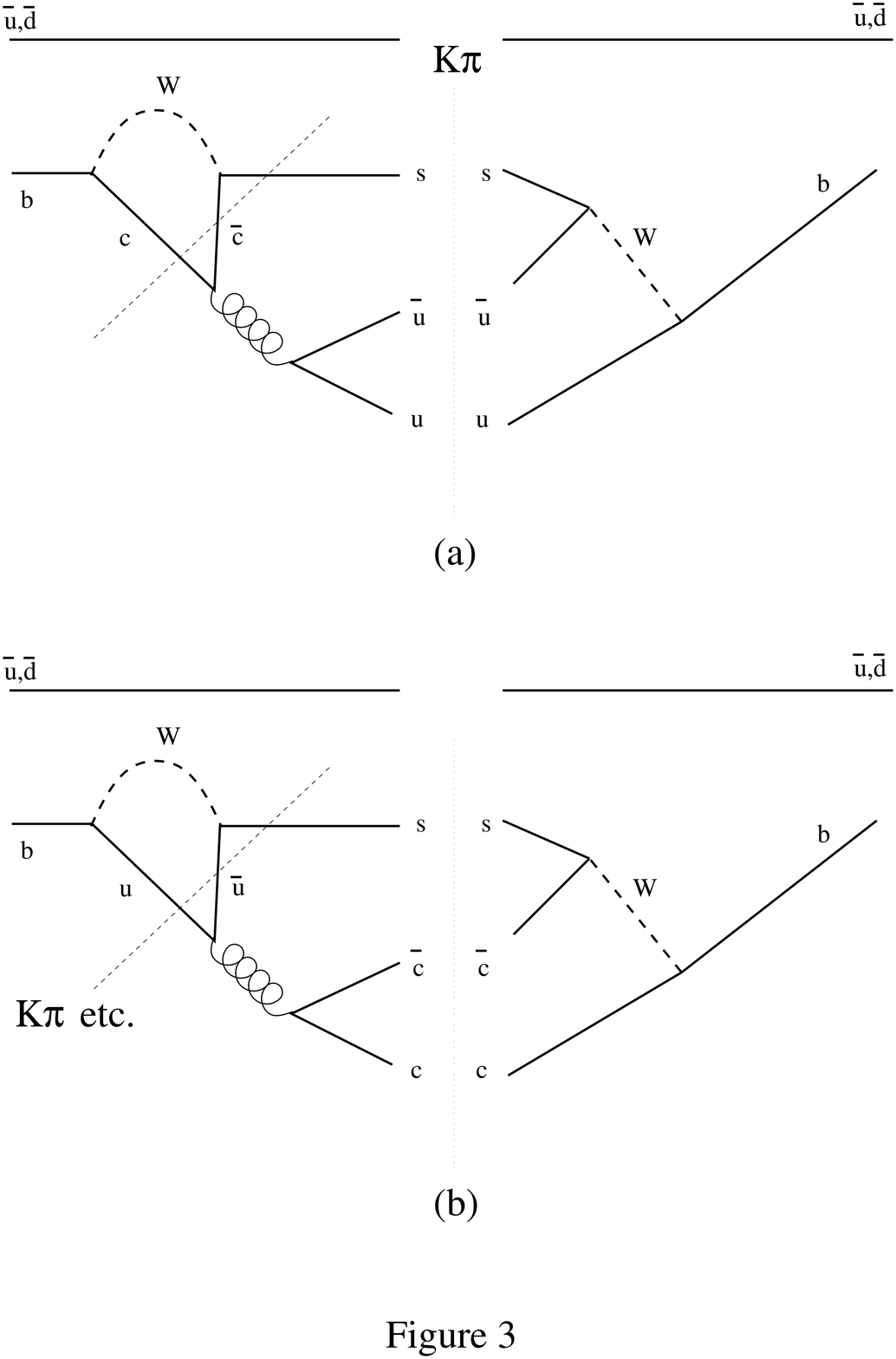}}
\end{figure}

\newpage

\begin{figure}[ht]
\hspace*{-0.5in}
\vspace*{0in}
\epsfysize 7 in
\mbox{\epsfbox{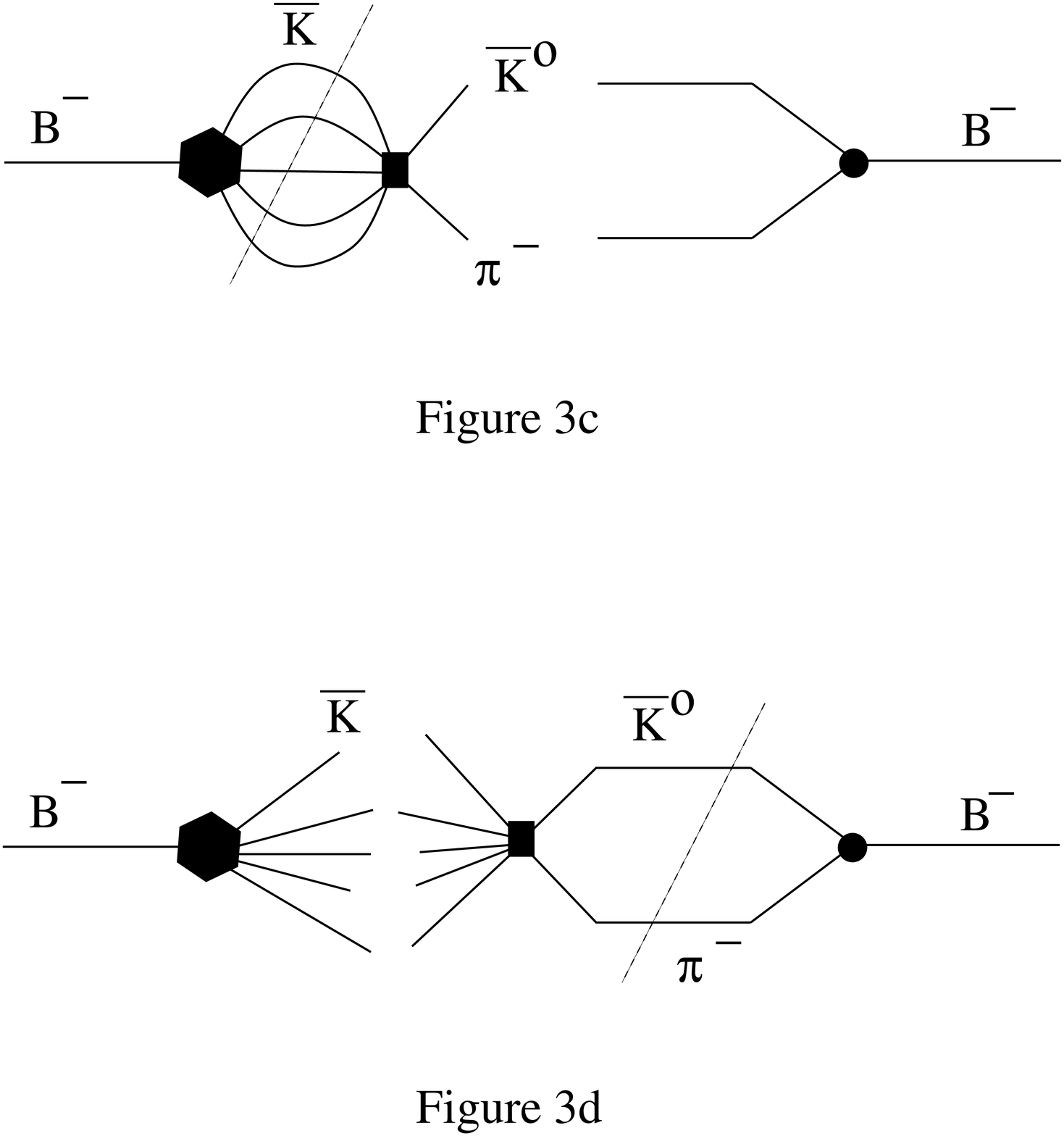}}
\end{figure}

\newpage

\begin{figure}[ht]
\hspace*{-0.5 in}
\vspace*{ 0 in}
\epsfysize 5 in
\mbox{\epsfbox{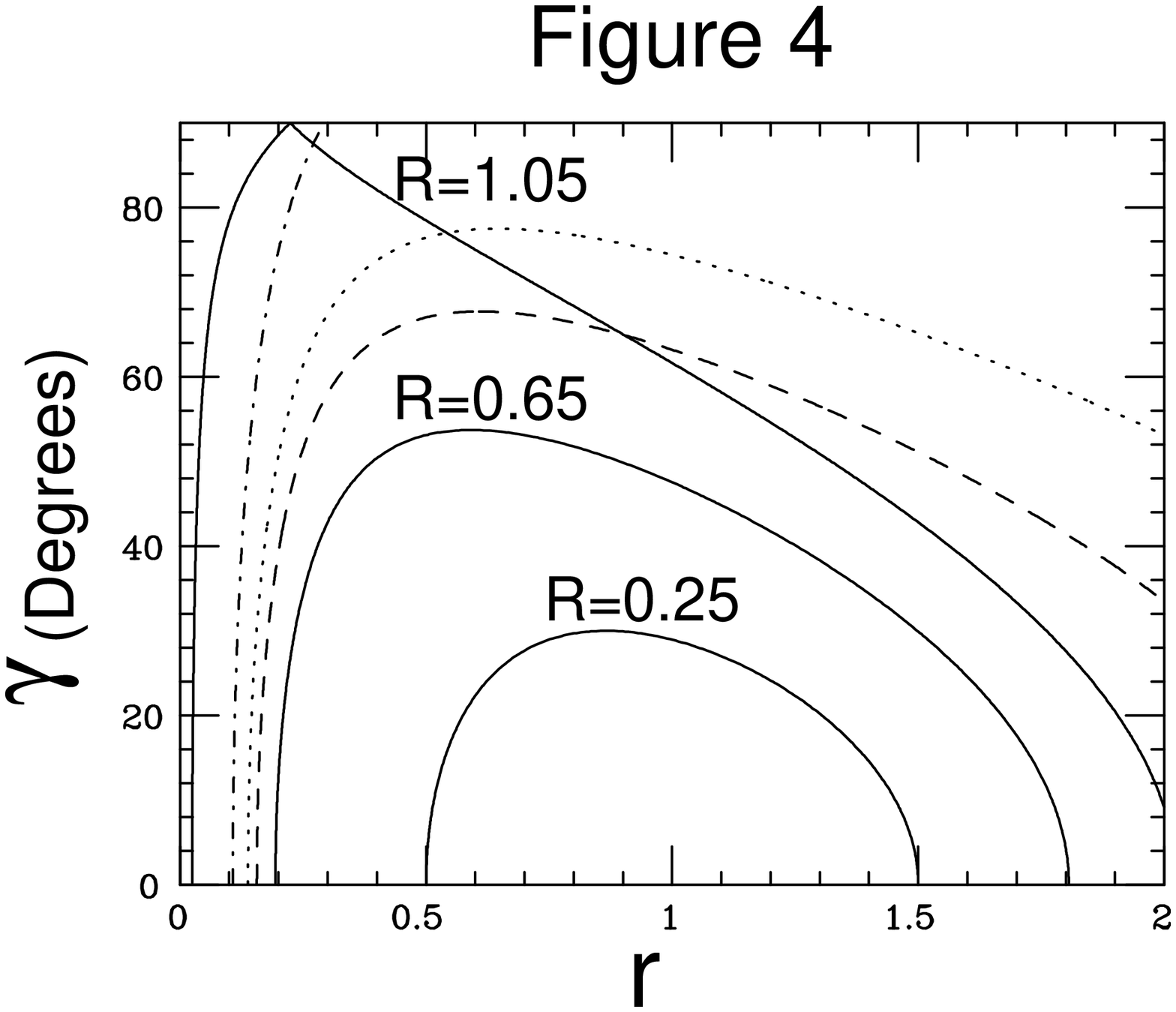}}
\end{figure}

\newpage

\begin{figure}[ht]
\hspace*{-0.5 in}
\vspace*{ 0 in}
\epsfysize 5 in
\mbox{\epsfbox{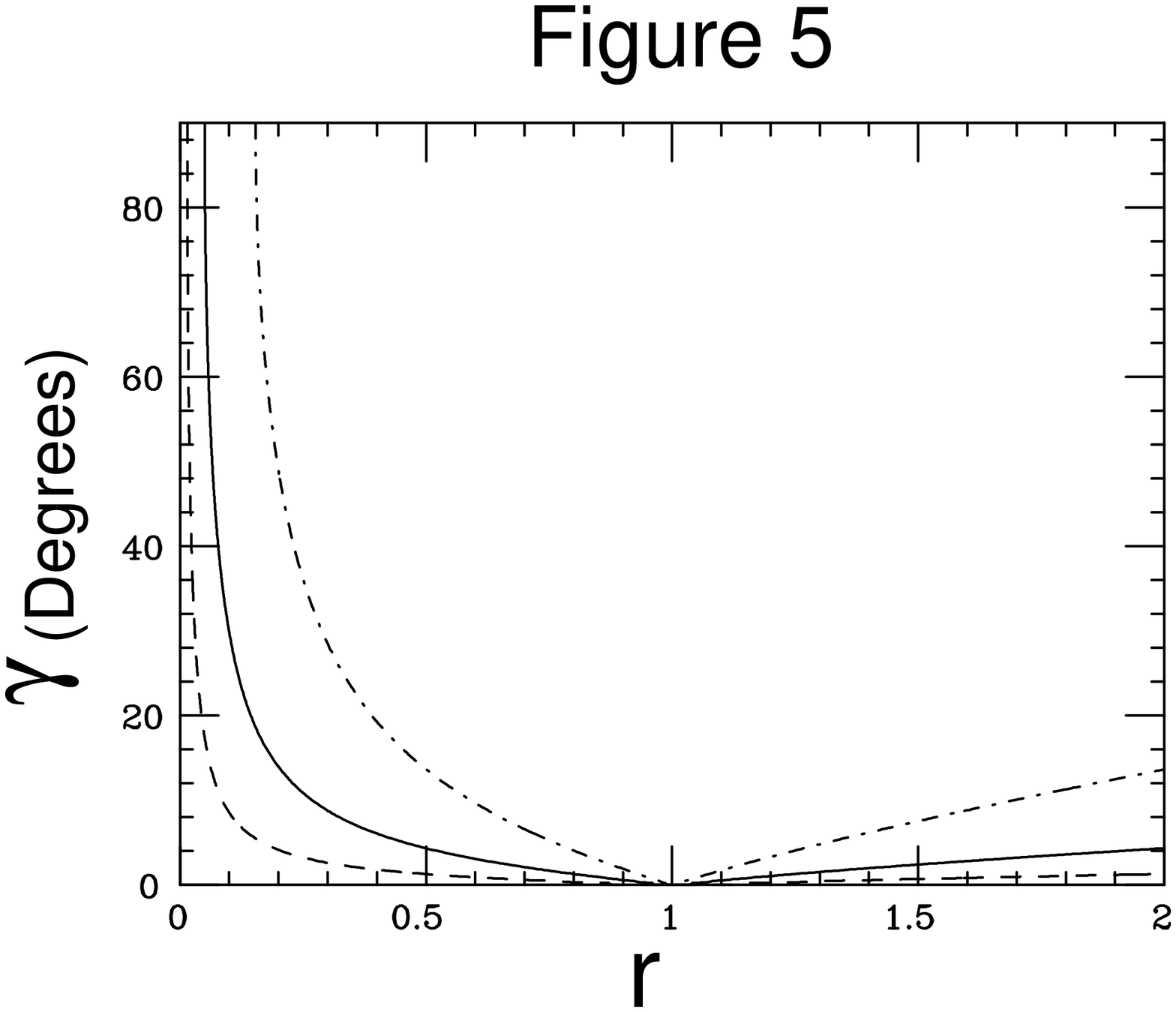}}
\end{figure}


\begin{thebibliography}{99}
\bibitem{weakex} These decays receive contributions from weak exchanges
involving the spectator quark (i.e. non-spectator decays). 
We take the view here that there is no unique
separation of such contributions from final state interactions of usual
spectator model
decays.
\bibitem{inelastic} The importance of the inelastic contribution to the
rescattering has also been emphasized by L. Wolfenstein~\cite{wolf2}.
\bibitem{notpra} Strictly speaking $\Delta\Gamma (A\to X)$ (See
eq.~(\ref{pradef})) is not PRA  but rather is
partial width difference between conjugate modes.
PRA is defined to be the ratio 
$\frac{\Gamma(A\to X)-\Gamma (\bar A\to \bar X)}{\Gamma (A\to X) +
\Gamma (\bar A\to \bar X)}$.  
We will often ignore this
distinction; the context should make it clear as to which is
appropriate.
\bibitem{isospin} Note that, in the isospin expansion, $B$ and $C$ need
not be equal. This is because in tree decays of $B^\pm$ there are two
$\bar u$ quarks in the final state which may be interchanged while in
$\bar B^0$ decay all the final states are distinct.
\end{thebibliography}
\end{document}